\documentclass[%
 reprint,superscriptaddress,
 amsmath,amssymb,pra]{revtex4-1}

\usepackage{physics}
\usepackage{xcolor}
\usepackage{xtab,afterpage,longtable}
\usepackage{lipsum}
\usepackage{graphicx}
\usepackage{dcolumn}
\usepackage{booktabs} 
\usepackage{multirow}
\usepackage{color}
\newcommand*{\rom}[1]{\expandafter\@slowromancap\romannumeral #1@}
\usepackage{hyperref}
\usepackage{orcidlink}

\usepackage[sectionbib]{bibunits}
\defaultbibliographystyle{apsrev4-2} 
\defaultbibliography{bibliography}

\begin{document}
\begin{bibunit}[apsrev4-2]

\title{Quantum sensing with arbitrary frequency resolution via correlation measurements}

\author{Jungbae Yoon\,\orcidlink{0009-0000-9591-243X}}
\affiliation{
  Research Laboratory of Electronics, Massachusetts Institute of Technology, Cambridge, MA 02139, USA}
\affiliation{Department of Physics, Korea University, Seoul, 02841, Republic of Korea}

\author{Keyuan Zhong}
\affiliation{
  Research Laboratory of Electronics, Massachusetts Institute of Technology, Cambridge, MA 02139, USA}
\affiliation{
  Department of Nuclear Science and Engineering, Massachusetts Institute of Technology, Cambridge, MA 02139, USA}

\author{Guoqing Wang\,\orcidlink{0000-0002-1822-8121}}
\affiliation{
  Research Laboratory of Electronics, Massachusetts Institute of Technology, Cambridge, MA 02139, USA}

\author{Boning Li\,\orcidlink{0000-0003-3979-0778}}
\affiliation{
  Research Laboratory of Electronics, Massachusetts Institute of Technology, Cambridge, MA 02139, USA}
\affiliation{Department of Physics, Massachusetts Institute of Technology, Cambridge, MA 02139, USA}

\author{Donghun Lee\,\orcidlink{0000-0003-3348-9474}}
\affiliation{Department of Physics, Korea University, Seoul, 02841, Republic of Korea}

\author{Paola Cappellaro\,\orcidlink{0000-0003-3207-594X}}%
\email{pcappell@mit.edu}
\affiliation{
  Research Laboratory of Electronics, Massachusetts Institute of Technology, Cambridge, MA 02139, USA}
\affiliation{
  Department of Nuclear Science and Engineering, Massachusetts Institute of Technology, Cambridge, MA 02139, USA}
\affiliation{Department of Physics, Massachusetts Institute of Technology, Cambridge, MA 02139, USA}
  
\date{\today}
       
\begin{abstract}
Achieving high-frequency spectral resolution with quantum sensors, while crucial in  fields ranging from physical to biological sciences, is challenging due to their finite coherence time. Here, we introduce a novel protocol that achieves this goal by measuring phase correlations of AC magnetic fields using ensembles of NV centers. Our method extends the sensing dynamic range to frequencies higher than the system's Rabi frequency while achieving arbitrary frequency resolution, limited only by the target field coherence time. Moreover, our approach operates more robustly with respect to the magnetic field's amplitude. Thanks to this robustness, our protocol allows the application of more $\pi$-pulses in pulse sequences such as CPMG, enabling the decoupling of a broader range of frequency noise. The higher harmonics generated in this process continue to act as a part of the signal, ultimately improving the frequency resolution. This method paves the way for achieving arbitrary frequency resolution with improved performances, making it highly versatile for quantum sensing applications across diverse scientific fields.

\end{abstract}

\maketitle
\section{Introduction} \label{sec:introduction}
The ultimate goal of quantum sensing is to achieve excellent sensitivity, high spatial, temporal, and spectral resolution. Improving frequency resolution is of particular interest in many applications: In gravitational wave detection, precise frequency measurements are needed to detect tiny spacetime distortions~\cite{ligo2011gravitational,bailes2021gravitational}. Nuclear magnetic resonance, e.g., needs to accurately discriminate very small chemical shifts~\cite{appelt2006chemical,wenchel2024super}. In material science and nanotechnology, high frequency resolution is critical to understand spin dynamics in nanomagnets and spintronic devices~\cite{vavnatka2021spin,an2024emergent}. 

Nitrogen Vacancy (NV) centers in diamond have emerged as powerful sensors for the last two applications. In conventional methods~\cite{staudacher2015probing,zaiser2016enhancing}, the NV frequency resolution is typically limited by their coherence time and spin relaxation time. Decoherence and relaxation can be improved via dynamical decoupling sequences~\cite{de2010universal,chen2023extending} or by engineering and controlling the spin bath~\cite{balasubramanian2009ultralong,bauch2018ultralong,herbschleb2019ultra}. Still other efforts have been focused on bypassing the limits imposed by coherence times. 
One strategy inspired by stimulated echo experiments~\cite{van1988optically} is to store the NV into a population state during the evolution of the target system~\cite{laraoui2013high,liu2022surface,jiang2023quantum}, between encoding and decoding periods. Mapping the NV state to a nuclear spin memory can further overcome the NV $T_1$ relaxation limit that still afflicts this method~\cite{pfender2017nonvolatile}. 
Other methods have tackled the goal of achieving spectral resolution not limited by the sensor relaxation time, by using a sequence of shorter, synchronized measurements~\cite{schmitt2017submillihertz,boss2017quantum,glenn2018high}. Extending the total measurement time, $T$, yields a frequency resolution $\delta \omega\sim 1/T$, provided the target AC field is coherent over $T$. 
Unfortunately, a few drawbacks of these protocols have emerged, limiting their application. 
To reliably extract the AC frequency, synchronized measurements rely on the small-angle approximation of the qubit phase acquired during each measurement~\cite{boss2017quantum,aslam2017nanoscale}. When this approximation is not valid (whether because the AC signal amplitude is large or the acquisition time is long) not only  the simple, Fourier-transform based data analysis yields poor signal-to-noise ratio, as it discards many components
of the signal, but the qubit signal might vanish altogether for some values of the AC signal magnitude~\cite{boss2017quantum,mizuno2020simultaneous}.
These drawbacks can be avoided by limiting the phase acquisition time. Not only this requires prior knowledge of the magnetic field amplitude. For (relatively) slow fields, this also entails using a small number of $\pi$-pulses in the decoupling sequence (since the pulse spacing needs to be close to the AC frequency). This forfeits the opportunity to decouple a broader range of noise frequencies, decreasing $T_2$ or distorting the qubit response. 

In this work, we introduce a novel technique for achieving arbitrary frequency resolution by measuring correlations between sequential measurements of NV center ensembles. Our approach exhibits enhanced robustness to a broad range of magnetic field strengths, distinguishing it from synchronized measurement protocols. Although correlations between two (or more) NV centers have been proposed to distinguish between global and local signals~\cite{rovny2022nanoscale}, here we utilize correlations between multiple measurements to achieve arbitrary frequency resolution. 
Our technique enables AC magnetic field measurements without prior knowledge of the AC initial phase and magnetic field strength over a broad range of AC frequencies, even exceeding the Rabi driving frequency. 
In fact, we find that considering the correlations instead of the Fourier transform of the synchronized measurements cancels the effects of random phases~\cite{meriles2010imaging,herbschleb2024robust} and avoids spreading the target signal information across higher harmonics of the Fourier transform~\cite{boss2017quantum,mizuno2020simultaneous}. Contrary to Fourier-based methods, we obtain a further frequency resolution improvement as the qubit acquired phase increases (either due to larger field magnitudes or longer phase acquisition times). As a result, we can increase the number of $\pi$-pulses, thereby enhancing the decoupling of a wider range of noise frequencies. We experimentally demonstrate that with a 2.5 second measurement time, we can resolve frequency differences as small as 0.2 Hz at 5 MHz. Finally, we show that utilizing ensemble NV centers along multiple axes can further improve the efficiency of our protocol by saving significant time with simultaneous initialization and readout of all NV centers and enhancing the contrast and frequency resolution compared to measurements conducted with fewer axes.

\section{Protocol for Correlation Sensing} \label{sec:results}

Our protocol proceeds similarly to synchronized measurements (see Fig.~\ref{fig:2 orientation correlation}), where the effect of an AC signal, $B(t)=B\cos(\omega t+\phi_0)$, on the evolution of the spin-qubit coherence is measured at fixed intervals, $t_d=i(\tau+t_{dead})$, where $\tau$ is the spin qubit phase acquisition time and $t_{dead}$ a dead time in between acquisitions (used, e.g., for initialization and readout.) Instead of performing the Fourier transform of the (average) signal, we analyze the correlations over the random phase $\phi_0$. In the simplest case of two measurement times yielding the signals $S_1, S_2$, the covariance $\text{Cov}(S_1, S_2)$ 
varies as a function of $t_{d}$~\cite{rovny2022nanoscale,SOM}, thus allowing extraction of the angular frequency of the AC signal, $\omega$. 

More generally, for $N_s$ time delays, we consider the sum over all correlations,
\begin{equation}\label{eqn:nsensors_0}
   \frac{1}{N_s^2} \sum_{i=1}^{N_s} \sum_{j=1}^{N_s} \text{Cov}(S_i, S_j)=\text{Var}\left( \frac{1}{N_s} \sum_{i=1}^{N_s} S_i \right),
\end{equation}
which can be extracted from the variance of the total signal. While the variance of individual signals, $\text{Var}(S_i)$ does not vary with $\omega$ and only adds to the background noise, measuring the variance of the total signal makes it possible to implement our method in situations where independent spin state readouts are challenging, such as in ensemble measurements or measurements of multiple single qubits within the diffraction limit.

\subsection{Correlation Measurement Using Ensemble NV Centers} \label{sec:correlation2}

\begin{figure*}[t]
\centering
\includegraphics[width=0.9\linewidth]{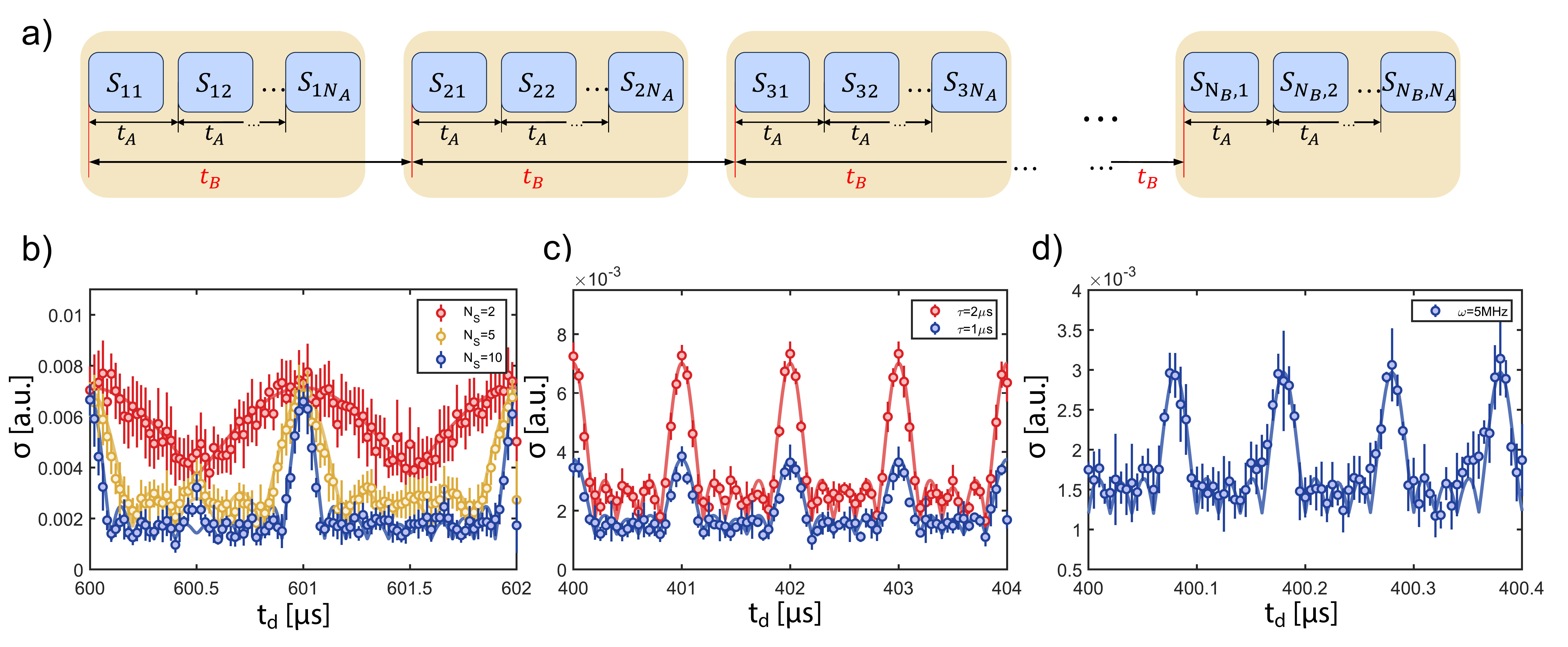}
\caption{\textbf{Correlation measurements using ensemble NV centers} (a) Correlation measurement protocol. (b) Linewidth reduction with increasing number of measurements ($N_s$). We measured a $\omega=500$~kHz AC magnetic field with 10 different phases ($N_\phi=10$) and 10 repetitions ($N_r=10$), setting $t_A=t_{d}$, $N_A=N_s$ and $N_B=N_\phi$. The frequency resolution improves as we go from $N_s=2$ (red), to $N_s=5$ (yellow), and $N_s=10$ (blue) correlations. Circles are experimental data with errorbars set by the standard deviation of the signal over $N_r$ repetitions. Lines are fits to Eq.~(\ref{eqn:nsensors}). Note that we normalized the variance in Eq.~(\ref{eqn:nsensors}) to the standard deviation, $\sigma$. (c) Correlation signal for different evolution times. The same AC signal, $\omega=500$~kHz, was measured with a $\tau=2\mu$s (red) and $\tau=1\mu$s (blue) echo time (including pulse durations). While the shorter time is only half the optimal echo time, the correlation is still measurable. Here $N_\phi=20$, $N_s=5$ and $N_r=5$. Circles are experimental data and lines fits to the square root of Eq.~(\ref{eqn:nsensors}). (d) Correlations measurements of a $\omega=5$~MHz signal, which is above our Rabi frequency ($\sim$3.2 MHz). Blue circles: experimental data. Lines, fit. We used $N_\phi=20$, $N_s=5$ and $N_r=5$.}
\label{fig:2 orientation correlation}
\end{figure*}

To measure the external magnetic field, one can apply an interferometric sequence to the spin qubit, utilizing an appropriate dynamical decoupling sequence to suppress noise. The accumulated phase, which contains information about the external magnetic field, can be estimated by measuring the population in $\ket0$, $S\propto P_{|0\rangle}$. When two measurements are performed with a time delay of $t_{d}$, they see an AC field with an additional phase difference given by $\omega t_{d}$. By averaging over the AC field's initial random phase $\phi_0$, the variance of the two measurements is
\begin{equation}\label{eqn:twosensors}
\begin{split}
&  \textrm{Var}\left(\frac{S_1+S_2}{2}\right)=\frac{1}{16}+\frac{1}{16}J_{0}[8\vartheta]-\frac{1}{4}J^{2}_{0}[4\vartheta]\\&+\frac{1}{16}\left(J_{0}[8\vartheta\sin{(\frac{\omega t_{d}}{2}})]+J_{0}[8\vartheta\cos{(\frac{\omega t_{d}}{2})}]\right),
\end{split}
\end{equation}
where $J_0$ is the zeroth-order Bessel function and $\vartheta$ is a constant that depends on the pulse sequence, the amplitude of the external magnetic field $B$, and the free evolution time $\tau$ (for spin echo, $\vartheta = \frac{\gamma_e B}{\omega} \sin^2\left(\frac{\omega \tau}{2}\right)$). As shown in Fig.~\ref{fig:2 orientation correlation}.b (red data), the two measurements exhibit correlations and anticorrelations as a function of $\omega t_{d}$ with period $\frac{\pi}{\omega}$ (note that this allows for better frequency resolution than the typical period of $\frac{2\pi}{\omega}$ obtained, e.g., from dynamical decoupling sequence with a final pulse that is $90^{\circ}$-phase shifted with respect to the initial pulse~\cite{rovny2022nanoscale}.) 

Instead of e.g., analyzing the Fourier transform for the two-measurement correlation at varying $t_d$, we consider the variance of $N_s$ measurements taken at intervals of $t_{d}$
\begin{equation}\label{eqn:nsensors}
\begin{split}
  V_{N_s}\equiv& \textrm{Var}\left(\frac{1}{N_s}\sum_{i=0}^{N_s-1} S_{i}\right)=-\frac{1}{4}J_{0}^{2}[4\vartheta]\\&+\frac{1}{8N_s^{2}}\sum_{i,j=0}^{N_s-1}
  J_{0}[8\vartheta\cos{(\frac{i-j}{2}\omega t_{d})}]\\&+\frac{1}{8N_s^{2}}\sum_{i,j=0}^{N_s-1}
  J_{0}[8\vartheta\sin{(\frac{i-j}{2}\omega t_{d})}].
\end{split}
\end{equation}
Similarly to what happens when going from two-slit interferences to $N$ slits, the variance for $N_s$ measurements yields narrower interference profiles ($\propto\frac{\sin^2(N_s\omega t_d)}{N_s^2\sin^2{(\omega t_d)}}$ for $\theta\lesssim 1$ ) that translate into improved frequency resolution (see Fig.~\ref{fig:2 orientation correlation}.b).

To improve the data acquisition efficiency and robustness, we can replace measuring over random AC phases with performing $N_{\phi}$ successive measurements with AC phases $\phi_i=\phi_0+i\frac{2\pi}{N_\phi}$, for $i=1,2,\dots,N_\phi-1$. Phase increments can be obtained by using appropriate time delays between acquisitions such that each one starts at a time $it_\phi$ with $\omega t_\phi=2\pi/N_\phi (\textrm{mod } 2\pi)$. Still, because these phases could have been taken as random, there is no need for prior knowledge of the initial phase $\phi_0$ of the AC magnetic field or for accurate setting of $t_\phi$.

To demonstrate our protocol, we performed measurements by adjusting two time delays, $t_A < t_B$, as shown in Fig.~\ref{fig:2 orientation correlation}.a. In the first method, we set $t_A\equiv t_{d}$ and $t_B\equiv t_\phi$. We perform $N_s$ measurements, sweeping $t_{d}$ for each $it_\phi$. The second method reverses this procedure: we first average over the phases, by setting $t_A\equiv t_\phi$ and $t_B\equiv t_{d}$. In both cases, the minimum value of the time delay, $t_{d}$, is limited by the duration of a single measurement ($\tau+t_{dead}$). Indeed, as discussed later, one can simply implement the protocol with a single series of consecutive measurements, later grouped as needed in $N_\phi\times N_s$ subsets. To achieve high frequency resolution, which  is set by $T=N_st$,  the second protocol is more efficient, since it performs $N_\phi$ measurements during the long $t_{d}$ dead time. Conversely, the first approach might bring benefits if the timing precision of the experimental apparatus degrades over $t_B$, since $t_\phi$ needs not be set very accurately.

As a first illustration of correlation measurements, we sweep the time delay, $t_{d}$, from 600 to 602 $\mu$s under a magnetic field with $\omega=500$~kHz, with $N_s= 2, 5$, and $10$, and $N_\phi=10$ (Fig.~\ref{fig:2 orientation correlation}.b). We repeat the measurement $N_r=10$ times, each with a random initial phase $\phi_0$, to confirm that results are independent of the initial phase. The experimental data is in good agreement with the theoretical prediction (Eq.~\ref{eqn:nsensors}) from random phase average. They further show the expected narrowing of the interference pattern as $N_s$ increase (and the presence of $N_s - 1$ minima.)

Additionally, even when the period of the AC magnetic field is not matched to the spin echo pulse's free evolution time (i.e. $\tau\neq 2\pi/\omega$), the magnetic field measurements remain feasible, albeit with some loss of signal contrast, since the maximum is $\overline V_{N_s}=\frac{1}{8} \left(1+J_0(8 \vartheta)-2 J_0(4 \vartheta)^2\right)$. This is shown in Fig.~\ref{fig:2 orientation correlation}.c, where we compare the signal for the optimal $\tau=2\mu$s echo time (red line) and for half time, $\tau=1\mu$s (blue line).

In Fig.~\ref{fig:2 orientation correlation}.d we further show that our protocol can measure an AC magnetic field at a frequency (of 5 MHz), which exceeds the Rabi frequency ($\sim$3.2 MHz in our setup). In this scenario, due to the finite pulse width, it is not possible to achieve the optimal free evolution time that matches the AC frequency. Still, sweeping $t_{d}$ enables measuring high-frequency signals. In previous covariance measurements using NV centers along different axes~\cite{rovny2022nanoscale}, differences in the NV coupling strengths to the target field, in their resonance and Rabi frequencies, contributed to further signal reduction that is not present here. Furthermore, when using NV centers along different axes, the projection of the magnetic field components results in a signal that is approximately halved, compared to when the B-field and NV center orientations are aligned, even for systems with the same signal-to-noise ratio (SNR). Additionally, when the Rabi frequency is not sufficiently fast, the frequency distortions of the measured AC magnetic field due to finite pulse width become a challenging issue to address precisely~\cite{ishikawa2018influence,yoon2023characterization}. However, by sweeping $t_{d}$ to measure phase correlation, accurate frequency measurements can still be achieved, regardless of pulse width effects.

\subsection{Arbitrary Frequency Resolution through Correlation Measurements} \label{sec:correlation3}

\begin{figure*}[t]
\centering
\includegraphics[width=\linewidth]{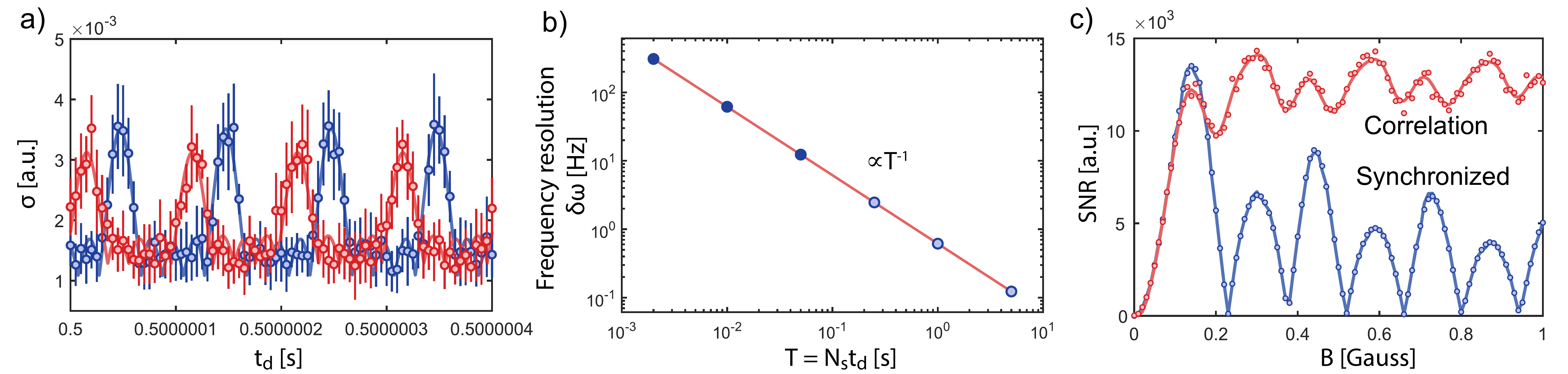}
\caption{\textbf{High-frequency resolution based on correlation measurements} (a) We demonstrate resolving   a 5MHz AC signal (blue) from a signal shifted by 0.2Hz (red). 
The fitting (line) to the experimental data (circles) yields a $0.31$Hz frequency difference between the two signals. This discrepancy is explained by the differences between the signal generator frequency and the pulse generator which operates with a different clock. (b) Frequency resolution dependence on total delay time, $N_s t_{d}$. We obtain the signal linewidth, $\delta t$ = full width at half maximum (FWHM) from the fit of the our measurement data (blue dots, $N_s=5,~\omega=5$ MHz) and evaluate the corresponding frequency resolution as $\delta\omega= \frac{\delta t}{t_d}\omega$. The red line is the frequency resolution derived from the linewidth of the square root of Eq.~(\ref{eqn:nsensors}). The first three points on the left correspond to our system's long measurement time, where $t_d = t_A$ was used in Fig.~\ref{fig:2 orientation correlation}.a. The last 3 points were acquired with $t_d=t_B$. Both protocols show a frequency resolution  inversely proportional to $N_s t_{d}$, similar to synchronized readout techniques. (c) Comparison of simulated SNR for correlation measurement (red) and synchronized readout (blue) as a function of the AC magnetic field strength $B$. The simulation data (dots) assumed a $\omega=500$~kHz magnetic field, a spin echo with $\tau=2~\mu$s and $t_{d}=2m~\mu s+2$ ns (m is an integer). We used $N_\phi=1000$ and $N_s=10$ for the correlation measurement and we also obtain the absolute square of DFT from the same consecutive measurements for the synchronized readout. Our simulations match well with the analytical results (lines, see~\cite{SOM}.)}
\label{fig:5MHz_compare}
\end{figure*}
Thanks to the interferences between the $N_s$ signals, the correlation Eq~(\ref{eqn:nsensors}) provides high frequency resolution. Indeed, for $\vartheta\lesssim 1$, $V_{N_s}\approx \overline V_{N_s}\frac{\sin^2(N_s\omega t_{d})}{N_s^2\sin^2(\omega t_{d})}$, which has the first zero at $\pi/(t_{d}N_s)$ (corresponding to the first uncorrelated phase difference.) As a result, the frequency resolution $\Delta \omega$ is inversely proportional to $t_{d} N_s$, which corresponds to the total measurement time when using the second protocol. 

Using this approach, we resolved two AC signals with $\omega$ centered around 5~MHz with a 0.2 Hz frequency difference in a 2.5~s total experimental time (see Fig.~\ref{fig:5MHz_compare}.a). Remarkably, this frequency resolution is possible even if the AC signal and our pulse sequence are not synchronized, since the AC signal and pulse generators operate with independent clocks. As a result, the actual frequency difference between the two AC signal is 0.31 Hz and they both have a 250 Hz offset. Still, we are able to achieve high frequency resolution even at high frequency. To evaluate the frequency resolution of our protocol, we plot the linewidth of our experimental correlation measurements as a function of the total measurement time, $N_st_{d}$, in Fig.~\ref{fig:5MHz_compare}.b, which clearly shows the expected behavior over a large range of times and frequencies.

Our correlation measurement protocol has several advantages with respect to other synchronized readout techniques that have been proposed in the past to achieve the same goal of arbitrary frequency resolution. Both our correlation measurement protocol and previous synchronized readout methods assume that the AC signal we wish to measure is highly coherent. In previous synchronized measurement protocols, the frequency information was typically extracted from the first harmonic of the signal's discrete Fourier transform (DFT). However, for some values of $\vartheta$ (that is, of the magnetic field and dynamical decoupling sequence settings) the first harmonic vanishes~\cite{boss2017quantum,mizuno2020simultaneous}. While it is still possible to extract the desired frequency information from higher harmonics~\cite{nishimura2022floquet}, the data analysis becomes cumbersome. Conversely, one would need prior knowledge of the magnetic field amplitude to determine the number of $\pi$-pulses in dynamical decoupling sequences~\cite{pham2012enhanced} required to avoid the point where the signal disappears. In contrast, correlation measurements can achieve high frequency resolution for any non-zero value of the magnetic field magnitude, which only determines the overall signal contrast. Indeed, all Fourier harmonics contribute to the maximum (constructive) correlation at $t_d=\pi/\omega$, while they interfere at other $t_d$ times, further narrowing the measured signal~\cite{SOM}. Being able to apply a large (optimal) number of $\pi$-pulses also allows for better decoupling of a broader range of low-frequency noise other than the desired frequency\cite{biercuk2011dynamical,degen2017quantum}. Therefore, in systems with noise at different frequencies, the correlation measurement is more powerful than the synchronized readout method, which is limited by the number of $\pi$-pulses depending on the amplitude of the magnetic field signal.

To compare the two different protocols, we calculated their respective SNR, as shown in Fig.~\ref{fig:5MHz_compare}.c. We assumed measurements using a sufficient number of ensemble NV centers, such that quantum projection noise could be averaged out. Each measurement was assumed to follow a Poisson distribution as arising from photon shot noise from an ensemble of NVs with a simplified model that assumes photons are collected only if the NVs are in $\ket{0}$. For the same number of measurements ($N_s=10$, $N_\phi=1000$), we observed how the SNR varied with magnetic field, finding that correlation measurements provide a robust SNR over a wide range (details of the calculations and simulations can be found in the supplementary material~\cite{SOM}). Additionally, we find that the signal displays narrower linewidth than expected as $B$ increases, enabling an enhanced frequency resolution, scaling as $\sim 1/B$~\cite{SOM}.

\subsection{Correlation Measurement of Ensemble NV Centers Using Different Orientations} \label{sec:correlation1}

\begin{figure}[t]
\centering
\includegraphics[width=\linewidth]{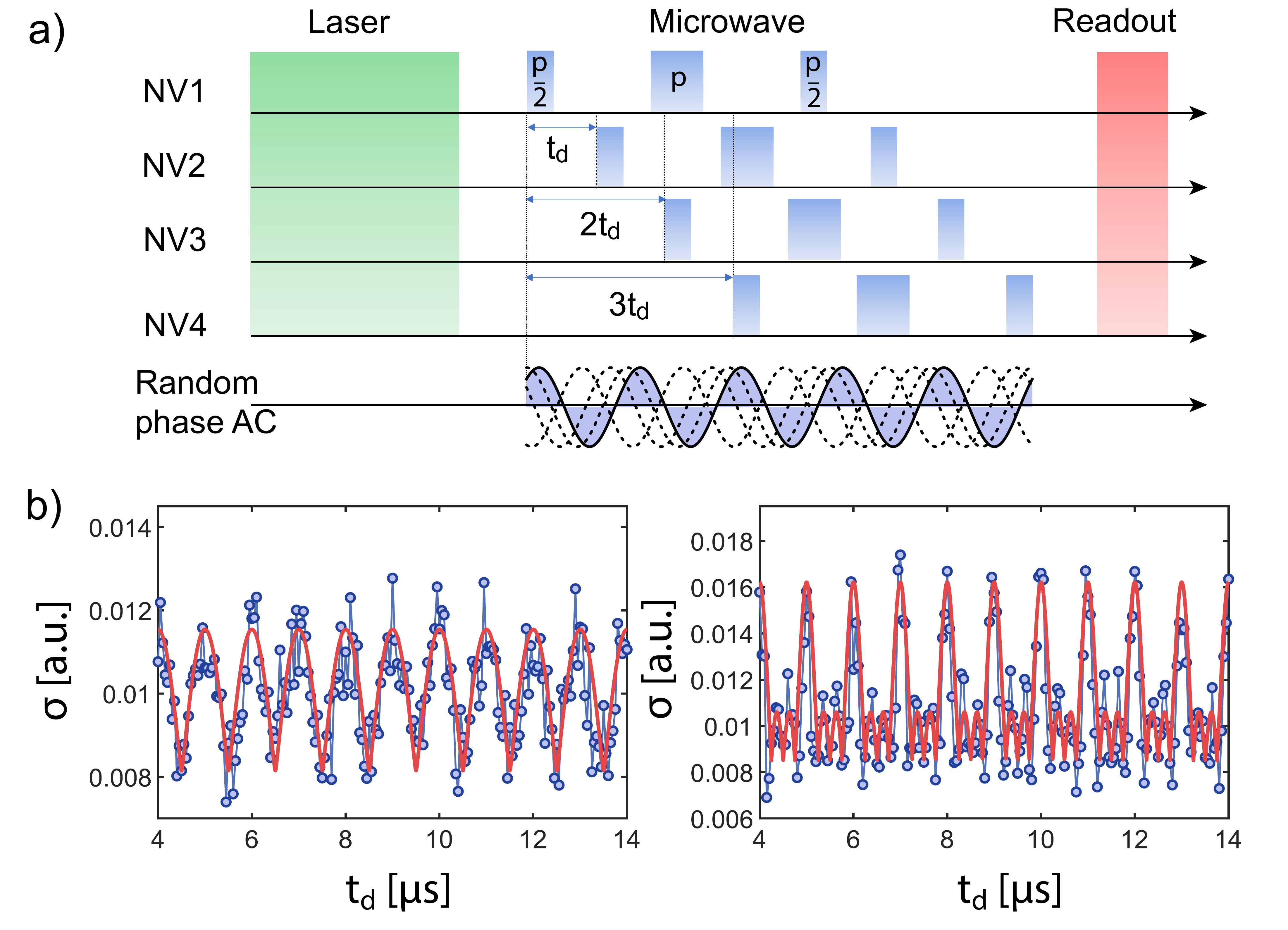}
\caption{\textbf{Spin echo correlation using ensemble NV centers with different orientations.} (a) Pulse sequence for the correlation measurement of ensemble NV centers along four different orientations. (b) Experimental results (blue dots) and theoretical calculations (red lines) of correlation measurements for two ensembles (left) and four ensembles (right) under a 500 kHz AC magnetic field, with a varying time delay $t_{d}$.}
\label{fig:4_orientation_correlation}
\end{figure}
While so far our method has exploited a measurement protocol similar to synchronized measurement, correlation measurement can be extended to other scenarios. A modified protocol, exploiting NV centers with distinct frequency addressability, can be used to achieve greater flexibility and efficiency in the data acquisition. Consider $N_s$ (ensembles of) NV centers that have distinct resonance frequencies. This could be the $N_s=4$ classes of NV spins aligned along the 4 crystallographic axes or NV spins in the presence of a magnetic field gradient~\cite{arai2015fourier}. Each subset of NV spins is driven with a separate control sequence, starting with a time delay $t_{d}$ (see Fig.~\ref{fig:4_orientation_correlation}.a). While the optical readout simultaneously collects lights from all spins, the frequency addressability enables implementing the correlation measurement. Notably, the time delay $t_{d}$ can be smaller than the single measurement time, thus avoiding inefficiency related to the initialization and optical readout times. 

To demonstrate this protocol, we performed measurements at a magnetic field $\vec B=[2.1, 10.6, 19.5]$Gauss (where z is aligned with [111]) to ensure that the resonance frequencies of NV centers along different axes are not degenerate.
As shown in Fig.~\ref{fig:4_orientation_correlation}.b, we used NV centers along two (left) and four (right) different orientations to detect a 500 kHz magnetic field. Not only we achieved an enhancement in contrast when using four orientations compared to two orientations, but also the expected improvement in frequency resolution. Note that the contrast increase is lower than what ideally expected (a factor of two) since the NV four orientations exhibit different and reduced contrast and coherence times due to the magnetic field misalignment~\cite{stanwix2010coherence, tetienne2012magnetic}. 

This method can be used to measure an AC signal even if its coherence is shorter than the single measurement time, $\tau+t_{dead}$. Since correlation measurements are based on random phase sampling, the method enables frequency resolution measurements limited by the shortest of the NV center's spin–lattice relaxation time $T_1$ and the coherence of the AC signal.

\section{Discussion and Conclusion} \label{sec:discussion}

Correlation measurements are often associated with low signal-to-noise ratios, particularly in the presence of significant noise~\cite{piao2016quantifying,massaro2023correlated}. However, by employing an ensemble of NV centers and utilizing a protocol that averages over phases through the strategic use of time between delays, we demonstrate that this limitation can be effectively mitigated. In addition, our experiments demonstrate that by utilizing the temporal correlations of sequential measurements, arbitrary frequency resolution comparable to previous studies can be achieved using ensemble NV centers. Furthermore, the robustness to higher harmonics not only eliminates the need for prior knowledge of the magnetic field amplitude but enables even higher frequency resolution. This allows us to apply as many $\pi$-pulses as possible in the dynamical decoupling sequence, further improving the frequency resolution within the same measurement time and suppressing low frequency noise. Additionally, when the amplitude of the AC signal is sufficiently large~\cite{herbschleb2024robust}, we can achieve an arbitrary frequency resolution by switching the pulse sequence to a Ramsey sequence. This extends the dynamic range from low frequencies (almost DC) to GHz (limited by the pulse generator sampling rate).

The method can be implemented using a single NV center without the need to apply single-shot measurements, providing we use enough repetitions, $N_s$ to achieve good SNR. Indeed, the relevant quantity is not the individual data points $S_i$, but rather the averaged signal, $\frac{1}{N_s} \sum_{i=1}^{N_s} S_i$. In addition to allowing the multi-NV protocol with simultaneous optical readout, collecting just the variance of the total signal implies that it is not necessary to store each individual $S_i$. This approach significantly reduces both the memory requirements and computational time, offering practical improvements in efficiency e.g., for imaging with camera which requires large size of data.

Given these advantages and the demonstrated high frequency resolution, our method holds broad potential applications, including low magnetic field NMR with single-defect spatial resolution, and could enable high-frequency resolution experiments even in more general situations. As an outlook, we point out that our correlation method could even be applied to DC magnetic field sensing, by exploiting the NV coupling to nuclear spins (preferentially hydrogens due to their large gyromagnetic ratio.) Thanks to the frequency resolution of our method, one could detect static magnetic fields with a $T^{-1}$ precision, where $T$ is only limited by the \textit{nuclear} spin $T_2^*$ time. Despite the lower nuclear spin gyromagnetic ratio, taking advantage of the long nuclear spin dephasing time and the more favorable time scaling~\cite{schmitt2017submillihertz} ($T^{-1}$ vs. $T^{-1/2}$) than e.g., Ramsey protocols, can yield good magnetic field sensitivity. Additionally, since the frequency of local signals, such as the Larmor frequency, can slightly vary due to different local environments~\cite{jaskula2017superresolution}, performing high frequency resolution measurements on individual NV centers enables high spatial resolution detection of magnetic field variations.

\section*{Acknowledgments} \label{sec:Acknowledgements}

J.Y. thanks Junghyun Lee for fruitful discussions. This research was primarily supported by Cooperative Research on Quantum Technology (2022M3K4A1094777) through the National Research Foundation of Korea(NRF) funded by the Korean government (Ministry of Science and ICT(MSIT)). J.Y. acknowledges support from the education and training program of the Quantum Information Research Support Center, funded through the National Research Foundation of Korea (NRF) by the Ministry of Science and ICT (MSIT) of the Korean government (No. 2021M3H3A1036573).

\section*{Author contributions} \label{sec:contributions}

J.Y. conceived the correlation measurement technique and conducted the numerical simulations of SNR. J.Y. and P.C. developed the theoretical calculations. J.Y. and K.Z. performed the experiments with assistance from G.W. J.Y., B.L., and P.C. analyzed the data. All authors participated in the discussion. J.Y. and P.C. wrote the manuscript. P.C. supervised the overall project.

\section*{COMPETING INTERESTS} \label{sec:COMPETING INTERESTS}
The authors declare that they have no competing interests.

\putbib[bibliography]
\end{bibunit}

\pagebreak
\begin{bibunit}[apsrev4-2]

\widetext
\begin{center}
{\bf \large{Supplemental Materials:}}\\
{\bf \Large{Quantum sensing with arbitrary frequency resolution via correlation measurements}}
\end{center}

\setcounter{secnumdepth}{3}
\setcounter{equation}{0}
\setcounter{figure}{0}
\setcounter{table}{0}
\setcounter{page}{1}
\setcounter{section}{0}

\renewcommand{\bibnumfmt}[1]{[S#1]}
\renewcommand{\citenumfont}[1]{S#1}

\setcounter{figure}{0}
\renewcommand{\figurename}{FIG.}
\renewcommand{\thefigure}{S\arabic{figure}}

\setcounter{equation}{0}
\renewcommand{\theequation}{S\arabic{equation}}

\newcommand{\ham}{{\mathcal{H}}}
\newcommand{\sx}{\sigma_x}
\newcommand{\sy}{\sigma_y}
\newcommand{\sz}{\sigma_z}
\newcommand{\Red}[1]{\textcolor[rgb]{0,0,0}{#1}} 
\newcommand{\Blue}[1]{\textcolor{blue}{#1}} 
\newcommand{\pc}[1]{\textcolor[rgb]{.7,0,0}{#1}}
\newcommand{\gw}[1]{\textcolor[rgb]{.5,0,0}{#1}}

\section{Standard deviation signals readout methods}
\subsection{Standard Deviation of the Signal Due to Random Phases in Spin Echo Measurements}

The phase accumulated during the spin echo sequence is given by
\begin{equation}\label{echopopulation}
    \varphi=\gamma_eB\left[\int_{0}^{\frac{\tau}{2}}\sin{(\omega t+\phi_{0})}dt-\int_{\frac{\tau}{2}}^{\tau}\sin{(\omega t+\phi_{0})}dt\right]=-4\vartheta\cos\left(\frac{\omega\tau}{2}+\phi_{0}\right),
\end{equation}
with $\vartheta=\frac{\gamma_eB}{\omega}\sin^2(\omega\tau/2)$.
Here $\omega$ is the  frequency of the AC magnetic field, $\phi_{0}$ its phase  at the beginning of the spin echo pulse, $\tau$  the free evolution time, $B$  the amplitude of the magnetic field and $\gamma_e$ the electronic spin gyromagnetic ratio. We note that for different dynamical decoupling sequences we would find a similar expression but with a different value of $\vartheta$~\cite{degen2017quantum,ishikawa2018influence}. For example, the acquired phase for $n$ equidistant pulses is
\begin{equation}
    \label{eq:PDD}
    \varphi_{n,\text{PDD}}=-\frac{\gamma_eB}\omega \sin{\left(\frac{\omega\tau}2\right)}\tan{\left(\frac{\omega\tau}{4n}\right)}\cos{\left(\frac{\omega\tau}{2}+\phi_{0}\right)},
\end{equation}
\begin{equation}
    \label{eq:CP}
    \varphi_{n,\text{CP}}=-\frac{2\gamma_eB}{\omega} \sin{\left(\frac{\omega\tau}2\right)}\left(1-\sec{\left(\frac{\omega\tau}{2n}\right)}\right)\cos{\left(\frac{\omega\tau}{2}+\phi_{0}\right)},
\end{equation}
which can  in general be written as $\varphi_n\equiv\vartheta_n\cos{\left(\frac{\omega\tau}{2}+\phi_{0}\right)}$.

The probability of the qubit to be in the state $\ket0$ after the pulse sequence is then
\begin{equation}\label{echopopulation2}
    P_{\ket0}=\cos^{2}\left[\frac{\varphi}{2}\right]=\cos^{2}\left[2\vartheta\cos{\left(\frac{\omega\tau}{2}+\phi_{0}\right)}\right].
\end{equation}

If the pulse sequence is not synchronized with the AC signal (i.e., the AC signal has an initial random phase), the probability $P_{\ket0}$ averaged over the random phases becomes
\begin{equation}
    \mathbb{E}[P_{\ket0}]=\frac{1}{2\pi}\int_{-\pi}^{\pi}{\cos^{2}{[2\vartheta\cos{\left(\frac{\omega\tau}{2}+\phi_{0}\right)}]}}d\phi_{0}=\frac{1}{2}+\frac{1}{2}J_{0}[4\vartheta],
    \label{eq:mean}
\end{equation}
where $J_{0}$ is the zeroth-order Bessel function, while its variance is
\begin{equation}
    \text{Var}(P_{\ket0})=\frac{1}{8}+\frac{1}{8}J_{0}[8\vartheta]-\frac{1}{4}J_{0}^{2}[4\vartheta].
    \label{eq:var}
\end{equation}

\begin{figure}[t]
\centering
\includegraphics[width=\linewidth]{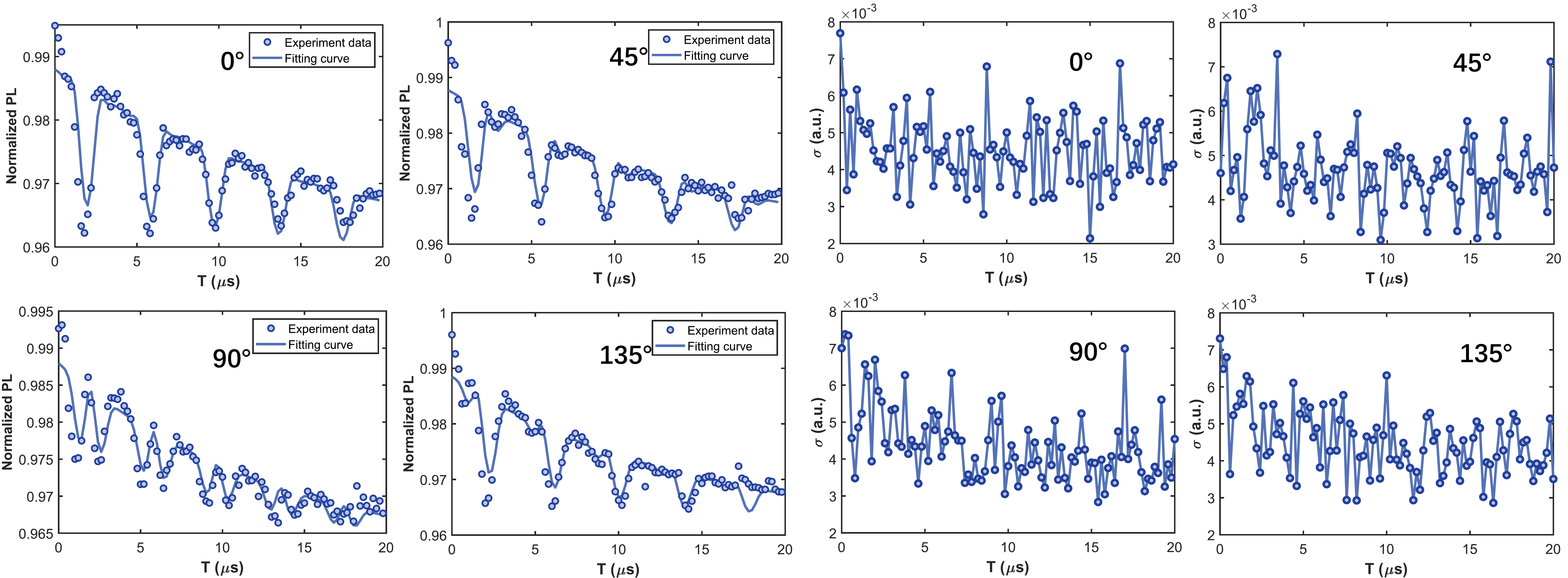}
\caption{\textbf{Spin echo measurement when our pulse is synchronized with AC magnetic field} \newline Spin echo measurement results as a function of the initial phase, $\phi_0$, of the 500 kHz magnetic field when synchronized. The mean value of 100 measurements (left) and the standard deviation (right).}
\label{fig:synchoronizedAC}
\end{figure}

Synchronizing the AC magnetic field phase and the spin echo pulse in the experiment results in identical measurements and a fixed $\phi_0=\varphi_0+n\omega t\equiv \varphi_0+n 2\pi$, thus leading to $\text{Var}(P_{\ket0})=0$. When instead the field period is not synchronized with the pulse sequence timing, $\phi_0$ varies, allowing to extract information about the field amplitude and/or its frequency from both the average and the variance of the signal, as shown in Fig.~\ref{fig:synchoronizedAC}.

\begin{figure}[t]
\centering
\includegraphics[width=\linewidth]{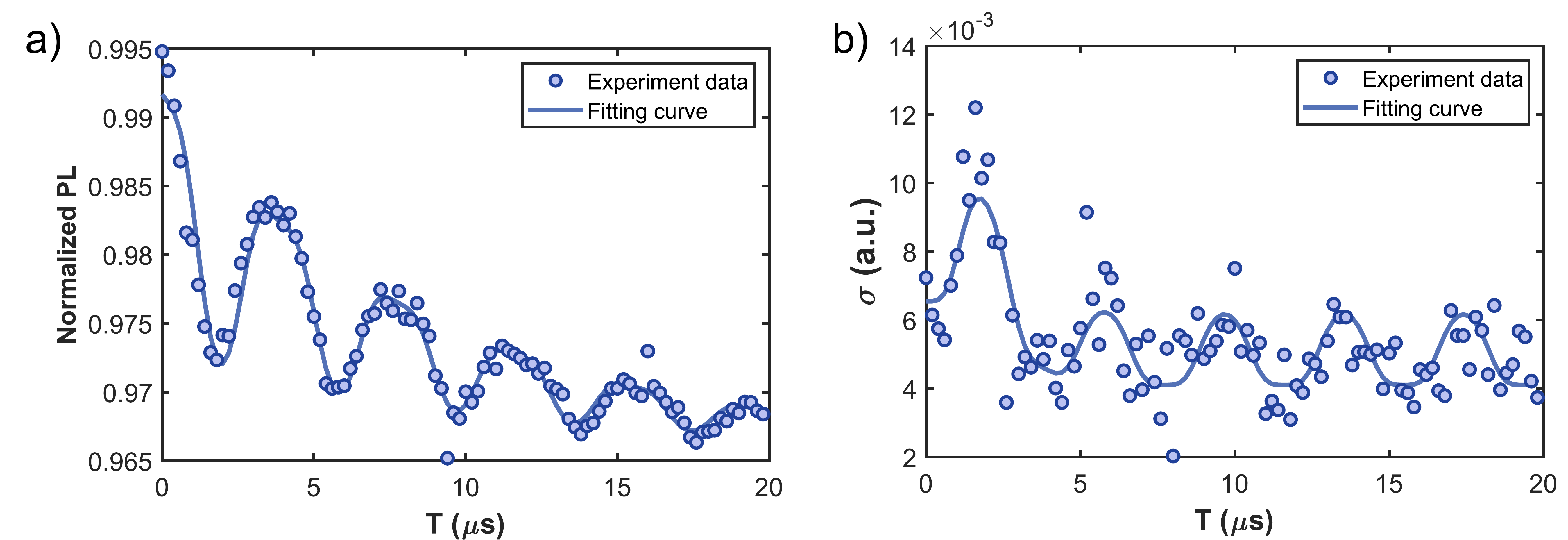}
\caption{\textbf{Spin echo measurement when our pulse is not synchronized with AC magnetic field} \newline Measurement results for the case where the initial phase of the 500 kHz magnetic field changes with each spin echo measurement when not synchronized: (a) mean value and (b) standard deviation}
\label{fig:nonsynchoronizedAC}
\end{figure}
However, we note that both average and variance depend only on $\vartheta$ (and thus $\tau$ but not t), which constraints the resolution with which one can resolve the frequency $\omega$. A similar limitation would be obtained even for synchronized measurement if the phase $\phi_0$ is unknown (or random over averaging of the experiments). To overcome this drawback, we propose to use correlations.

\subsubsection{Finite averages}\label{finite_avg}
In experiments, the number of phases $\phi_0$ over which we average to obtain the mean, variance (and correlations) is finite. In many cases, the phases are not truly random, but set by experimental delays (with at most some small random fluctuations around their nominal value). It thus becomes important to establish what is the minimum number of phases to be summed over.
As an example, we consider the mean (other averages would be similar since they involve the similar sinusoidal functions).
We replace Eq.~(\ref{eq:mean}) with a finite sum over phases: 
\begin{align}\label{finite_phases}
   \mathbf{E}_{N_\phi}[P_{\ket0}]=\frac{1}{N_\phi}\sum_{j=0}^{N_\phi-1}P_{\ket0}^j= \frac{1}{N_\phi}\sum_{j=0}^{N_\phi-1} \frac12\left(1+{\cos^{2}{[2\vartheta\cos{\left(\frac{\omega\tau}{2}+\phi_0+\frac{2\pi j}{N_\phi}\right)}]}}\right)
\end{align}

Given the expansion of exponentials in terms of Bessel function, $e^{i\theta\cos(\phi)}=\sum_{n=-\infty}^\infty i^ne^{in\phi}J_n(\theta)$, we can write the $j^{th}$ population as 
\begin{equation}
P_{\ket0}^j=\frac12[1+J_0(4\vartheta)]+\sum_{n=1}^\infty J_n(4\vartheta)\cos(n\pi/2)\cos[n(\omega t/2+\phi_0+\frac{2\pi j}{N_\phi})]
\label{Eq:SigBessel}
\end{equation}
We thus want to show under which condition the double sum
\begin{equation}
\frac{1}{N_\phi}\sum_{j=0}^{N_\phi-1}\sum_{n=1}^\infty J_n(4\vartheta)\cos(n\pi/2)\cos[n(\omega t/2+\phi_0)+nj\frac{2\pi }{N_\phi})]
\label{Eq:SigBessel0}
\end{equation}
is zero. We note that for $n\ll N_\phi$, $\sum_{j=0}^{N_\phi-1}\cos[n(\omega t/2+\phi_0)+nj\frac{2\pi}{N_\phi}]\approx 0$. 
For larger $n$ and small $\vartheta$, we instead have $J_n(4\vartheta)\approx (2\theta)^n/n!\approx 0$. We find thus that we need $N_\phi\gtrsim 2\theta$ for the discrete sum to approximate the (continuous) average. Similar results also apply when considering finite averages of variances and correlations. 

\subsection{Correlation of 2 Distinct Measurements}
We first consider acquiring the signal from either two NV centers or two subsequent measurements of the same NV. The two spin-echo pulse sequences are applied with a relative delay time $t_{d}$, which results in an additional phase shift $\phi_0'=\omega t_{d}$. Information about $\phi_0'$ is preserved in the correlation of the two measurements. However, to deal with the case where the readout happens simultaneously (as it would be case for measuring two NVs) we consider the combined population, 
\begin{equation}
    P_{\ket0}+P_{\ket0}^{2^{nd}}=\cos^{2}{[2\vartheta\cos{\left(\frac{\omega\tau}{2}+\phi_{0}\right)}]}+\cos^{2}{[2\vartheta\cos{\left(\frac{\omega\tau}{2}+\phi_{0}+\omega t_{d}\right)}]},
\end{equation}
and evaluate its variance. 

From  the average value of the signal, 
\begin{equation}
    \mathbb{E}[\frac{1}{2}P_{\ket0}+\frac{1}{2}P_{\ket0}^{2^{nd}}]=\frac{1}{2}+\frac{1}{2}J_{0}[4\vartheta],
\end{equation}
and the second moment,  $\mathbb{E}[\left(\frac{1}{2}P_{\ket0}+\frac{1}{2}P_{\ket0}^{2^{nd}}\right)^{2}]$, we obtain the variance
\begin{align}
   \textrm{Var}\left(\frac{1}{2}P_{\ket0}+\frac{1}{2}P_{\ket0}^{2^{nd}}\right)&\equiv \frac14\left[\textrm{Var}\left(P_{\ket0}\right)+\textrm{Var}\left(P_{\ket0}^{2^{nd}}\right)\right]+\frac12\textrm{Cov}\left(P_{\ket0},P_{\ket0,2nd}\right)\\
   &
=\frac12\left[\frac{1}{8}+\frac{1}{8}J_{0}[8\vartheta]-\frac{1}{4}J^{2}_{0}[4\vartheta]\right]+\frac{1}{16}\left[J_{0}[8\vartheta\sin\left(\frac{\omega t_{d}}{2}\right)] 
   +J_{0}[8\vartheta\cos{\left(\frac{\omega t_{d}}{2}\right)}]
   -2J^{2}_{0}[4\vartheta]\right].\nonumber
\end{align}

\subsection{Correlation of $N_s$ Distinct Measurements}
We can extend the result above to  $N_s$ distinct measurements, where  the $i$-th measurement ($i = 0, 1, 2, \dots, N_s-1$) is performed after a delay  $it_{d}$ relative to the $0$-th measurement. 
The normalized  signal from all $N_s$ measurements is the sum of the individual populations:
\begin{equation}\label{eq:nmeasurements0}
    \mathbb{P}=\frac1{N_s}\sum_{i=0}^{N_s-1} P_{\ket0 , i}=\frac1{N_s}\sum_{i=0}^{N_s-1}\cos^{2}{[2\vartheta\cos{\left(\frac{\omega\tau}{2}+\phi_{0}+i\omega t_{d}\right)}]}.
\end{equation}

The expectation value of the signal $\mathbb{P}$ over the random phase $\phi_0$ is then
\begin{equation}
    \mathbb{E}[\mathbb{P}]=\frac{1}{2}+\frac{1}{2}J_{0}[4\vartheta].
\end{equation}
We note that quite generally, for a large enough $N_s$ and non-synchronized acquisition, we have $\mathbb{P}\approx\frac12(1+J_0[4\vartheta])$.

By calculating the total signal second moment, 
\begin{equation}
\begin{split}
   \mathbb{E}[\left(\frac{1}{N_s}\sum_{i=0}^{N_s-1} P_{\ket0 , i}\right)^{2}]=\frac{1}{4}+\frac{1}{2}J_{0}[4\vartheta]+\frac{1}{8N_s^{2}}\sum_{i=0}^{N_s-1}\sum_{j=0}^{N_s-1}(J_{0}[8\vartheta\sin{\left(\frac{i-j}{2}\omega t_{d}\right)}]+J_{0}[8\vartheta\cos{\left(\frac{i-j}{2}\omega t_{d}\right)}]),
\end{split}
\end{equation}
we obtain the variance of the signal: 
\begin{align}\label{n_sensor correlations}
   \textrm{Var}\left(\frac{1}{N_s}\sum_{i=0}^{N_s-1} P_{\ket0}\right)&=\frac{1}{8N_s^{2}}\sum_{i=0}^{N_s-1}\sum_{j=0}^{N_s-1}\left[J_{0}[8\vartheta\sin{\left(\frac{i-j}{2}\omega t_{d}\right)}]+J_{0}[8\vartheta\cos{\left(\frac{i-j}{2}\omega t_{d}\right)}]\right]-\frac{1}{4}J_{0}^{2}[4\vartheta]\nonumber
   \\&=\frac{1}{8N_s}\left[1+J_{0}(8\vartheta)-2J^{2}_{0}(4\vartheta)\right]\\& -\frac{N_s-1}{4N_s}J_{0}^{2}(4\vartheta)+\frac{2}{8N_s^2}\sum_{i=1}^{N_s-1} (N_s-i)\left[J_{0}[8\vartheta\sin{\left(\frac{i\omega t_{d}}{2}\right)}]+J_{0}[8\vartheta\cos{\left(\frac{i\omega t_{d}}{2}\right)}]\right]\nonumber
\end{align}

We further note that for $\vartheta\leq 1$, we can approximate $J_{0}[8\vartheta\sin{\left(\frac{i\omega t_{d}}{2}\right)}]+J_{0}[8\vartheta\cos{\left(\frac{i\omega t_{d}}{2}\right)}]\approx A(\vartheta)+B(\vartheta)\cos(2i\omega t_{d})$. This makes it more apparent the analogy of our signal with the interference arising from $N_s$ slits. We thus find that 
\begin{align}
    \textrm{Var}\left(\frac{1}{N_s}\sum_{i=0}^{N_s-1} P_{\ket0}\right)&\underset{\vartheta\leq1}{\approx} \frac{1}{8}\left[1+J_{0}(8\vartheta)-2J^{2}_{0}(4\vartheta)\right]\frac{\sin^2(N_s\omega t_{d})}{N_s^2\sin^2(\omega t_{d})}
\end{align}
\section{Performance of Correlation Measurements}
\subsection{Single Time Measurement Performance in Ensemble of NV Centers}

NV centers spin readout at room temperature is hindered by two main issues: low collection efficiency of the emitted photons and NV reinitialization to its $\ket0$ state by the laser used for detection. As a result, obtaining information about the qubit state in a single measurement is not feasible. Instead, multiple repeated measurements are performed to obtain an average photon count. 

Using an ensemble of qubits is an effective solution to overcome the issue of low readout efficiency when the desired signal is the qubit expectation value. A sufficiently large ensemble of NV centers can provide a meaningful signal  in a single measurement despite the low collection efficiency, without requiring more  sophisticated techniques such as nuclear spin-enabled repeated readouts\cite{neumann2010single} or spin-to-charge conversion\cite{shields2015efficient}. 

\begin{figure}[t]
\centering
\includegraphics[width=\linewidth]{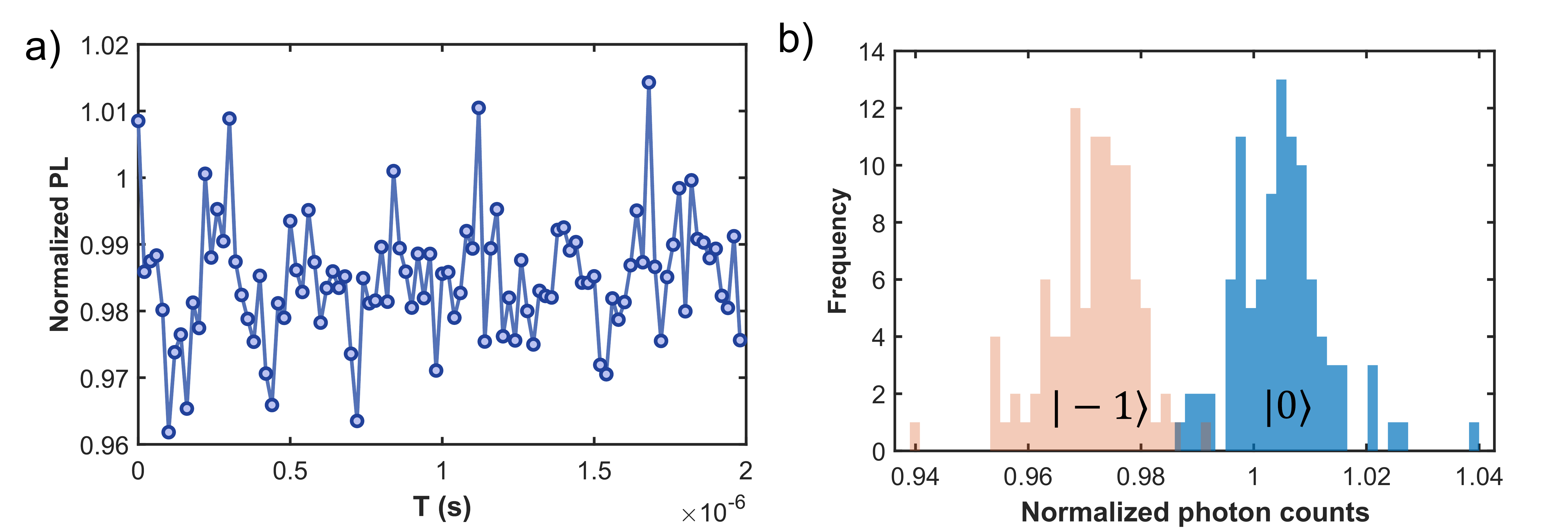}
\caption{\textbf{Performance of a single measurement in ensemble NV centers measurements} \newline (a) Rabi oscillation in a single measurement. (b) The normalized photon counts distribution of $\ket{0}$ and $\ket{-1}$ in a single measurement.}
\label{fig:Singleshot}
\end{figure}

We evaluate the performance of our readout in Fig.~\ref{fig:Singleshot}. The number of NVs in our ensemble and the setup collection efficiency are enough to detect Rabi oscillations in a single repetition of the experiment (Fig.~\ref{fig:Singleshot}.a). We can further evaluate the fidelity of the measurement by plotting the fluorescence measured after initializing the NV in their $\ket0$ state, and after an additional $\pi$ pulse prepares the $\ket{-1}$ states. Both measurements are normalized by the signal acquired by measuring $\ket0$ again, which is used as a reference signal to filter out effects from the large instability of our laser. The histogram of the measurement outcomes in Fig.~\ref{fig:Singleshot}.b clearly show that we can distinguish the two states. 

The data in Fig.~\ref{fig:Singleshot}.b does not follow a perfect Gaussian distribution, which can be attributed to the effects of the laser. Still, we can extract the means, $\alpha_{0,1}$ and variances, $\sigma_{0,1}$ from the histogram and calculate the contribution of the photon collection and noise to the signal variance as~\cite{rovny2022nanoscale} 

\begin{equation}\label{eqn:fidel}
\sigma_{R}^{gen}=\sqrt{1+2\frac{\sigma_{0}^{2}+\sigma_{1}^{2}}{(\alpha_{0}-\alpha_{1})^{2}}}.
\end{equation}
The measured value of 1.11 indicates a high fidelity of our measurement.

\subsection{Harmonics}
The phase a qubit acquires in AC field measurements varies with the field's initial phase (Eq.~(\ref{echopopulation2})). The maximum of the accumulated phase, $\varphi_{max}$, is proportional to $B$ (for stable $B$). Dynamical decoupling sequences, such as CPMG and XY8 with $N$ $\pi$-pulses, further increase the phase to $\varphi_{max} \propto B N$~\cite{degen2017quantum,boss2017quantum}.

In this case, the coherence time, $T_2$, and thus the interrogation time, $\tau$ can be extended with an increasing number of $\pi$-pulses, though not in a linear manner~\cite{bar2013solid}. Here, the number of $\pi$-pulses that can be applied is limited by both the coherence time of the NV center and experimental constraints such as finite pulse length and pulse errors. Since the pulse spacing is set by the target AC frequency, a longer $\tau$ allows for the application of more $\pi$-pulses, which might further increase $T_2$ by decoupling a broader range of low-frequency noise components. This makes the method particularly suitable when low-frequency noise is dominant~\cite{biercuk2011dynamical}. Ideally, to maximize the phase amplitude during magnetic field measurements, the time spacing between $\pi$-pulses should be maintained at half the target frequency period.

Therefore, considering these factors, one can measure by increasing the number of $\pi$-pulses in CPMG or XY8 to their limits. While usually beneficial, a large number of $\pi$-pulses might make the acquired phase $\varphi$ too large to take the usual approximation of the signal, $S\approx \varphi$. 
Then, we need to consider the effects of higher harmonics that arise from $\cos^2(\varphi/2)$ on both synchronized readout and correlation measurements.

\subsubsection{Synchronized readout}

\begin{figure}[t]
\centering
\includegraphics[width=\linewidth]{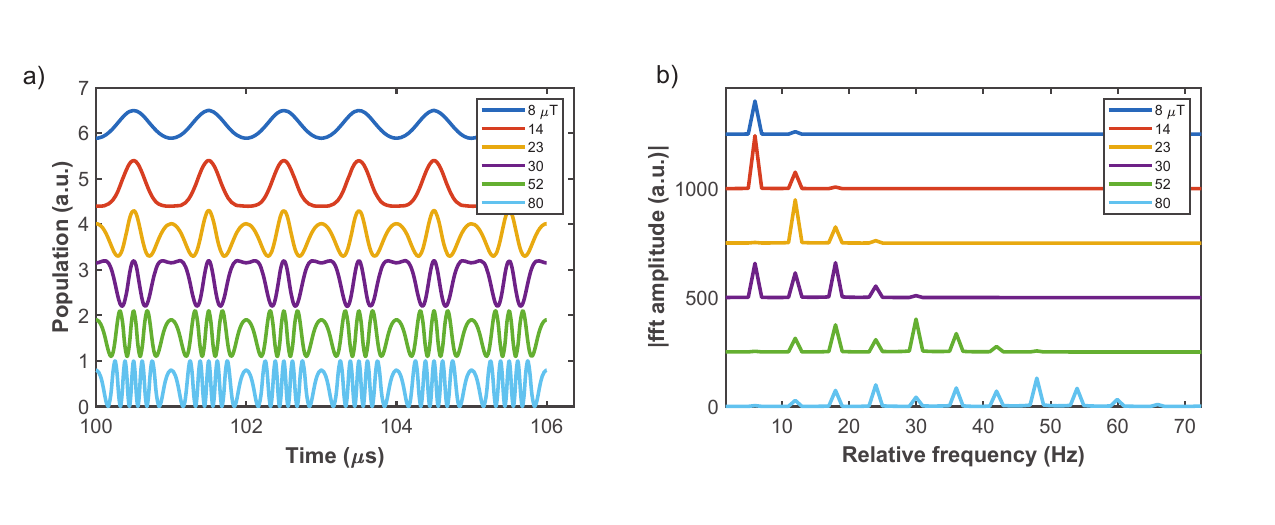}
\caption{ \textbf{Left: Population of the NV center as a function of time delay $t_d$ for different magnetic field amplitudes} as in Eq.~(\ref{echopopulation2}) with $\phi_0=\omega t_d$). We assume  a 500 kHz AC magnetic field with amplitudes of 8, 14, 23, 30, 52, and 80 $\mu$T.  \textbf{Right: Fourier transform showing increasing harmonics} Fourier transform of the left signal at a sampling frequency of 999.994 Hz.}
\label{fig:FFT_initial_phase}
\end{figure}

Fig.~\ref{fig:FFT_initial_phase}.a shows the qubit signal as a function of the AC phase, $\phi_0 = \omega t_d$, for different $B$ field amplitudes. Similar  behavior is also observed when increasing the number of $\pi$-pulses in the dynamical decoupling sequence at a fixed $B$ field amplitude, since both the $B$ field amplitude and the number of $\pi$-pulses contribute to increasing $\varphi_{max}$. 
This simulated data can be analyzed  via a Fourier transform (Fig.~\ref{fig:FFT_initial_phase}.b). 

Note that for certain magnetic field strengths, such as 23, 52, and 80 $\mu$T, the first harmonic almost disappears. While other harmonics can still provide frequency information~\cite{nishimura2022floquet}, the data analysis is more cumbersome and it would become complicated in the presence of multiple frequencies in the AC field~\cite{boss2017quantum}.

\subsubsection{Correlation measurement}

\begin{figure}[t]
\centering
\includegraphics[width=\linewidth]{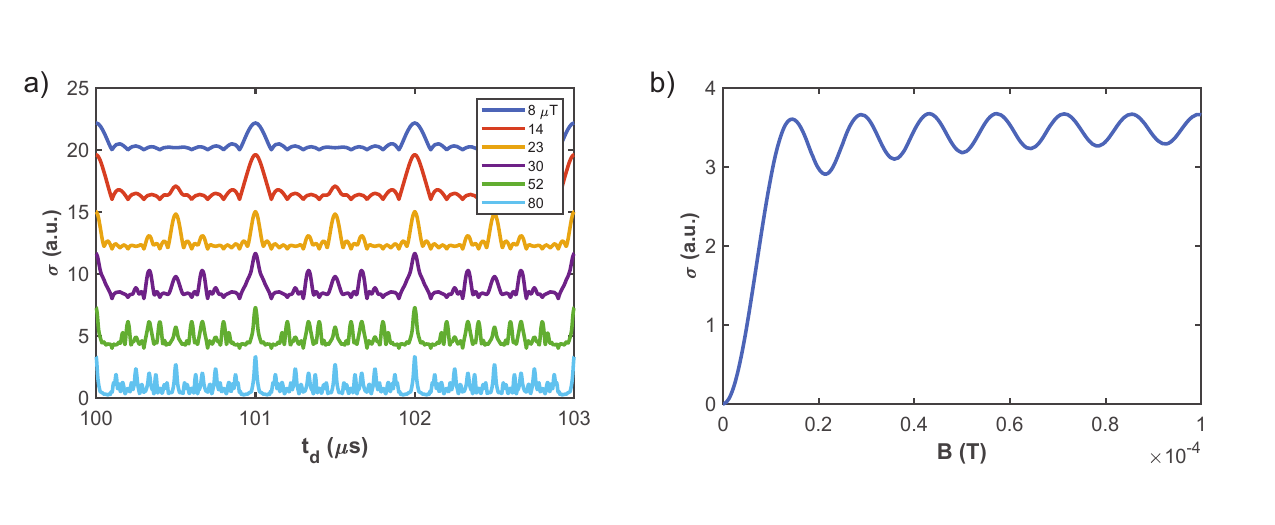}
\caption{\textbf{Example of the influence of higher harmonics on correlation measurements} \newline (a) We plot square root of Eq.~(\ref{n_sensor correlations}) for a measurement with $\omega_0=500$ kHz as a function of time delay $t_d$ for different magnetic field amplitudes. The parameters used here are $N_s = 10$. (b) Correlation signal at fixed time delay, $t_d=100$ $\mu s$ in (a) as a function of magnetic field amplitudes.}
\label{fig:correlation_measurement_harmonics}
\end{figure}

Higher harmonics are observed in correlation measurement as well (Fig.~\ref{fig:correlation_measurement_harmonics}). However, since higher harmonics generate interference patterns with frequencies that are integer multiples of the 1st harmonic frequency $\omega_0$, all higher harmonics peak at the same positions as the 1st harmonic. This phenomenon contributes to the robustness of our correlation measurement protocol with respect to magnetic field amplitude. 

\subsection{Frequency Resolution Improvement in Correlation Measurement}\label{sec:linewidth_narrow}

In general, the magnetic field strength $B$ that we wish to measure is fixed, and we can modify the phase amplitude by adjusting the number of $\pi$-pulses in spin echo, CPMG, or XY8 pulse sequences. When using CP-type pulse sequences for AC magnetic field measurements, the evolution time between $\pi$-pulses is usually optimized to match the target frequency $\omega$. Therefore, additional $\pi$-pulses can be added as long as the coherence time $T_2$ allows. If the magnetic field amplitude $B$ is small enough to avoid higher harmonics in the Fourier transform, the frequency resolution of both correlation measurements and synchronized readout is given by $\sim1/T$, where $T$ is the total measurement time. In the opposite regime, while higher harmonics are detrimental to Fourier-based synchronized readout, they can help correlation measurements.

\begin{figure}[t]
\centering
\includegraphics[width=0.6\linewidth]{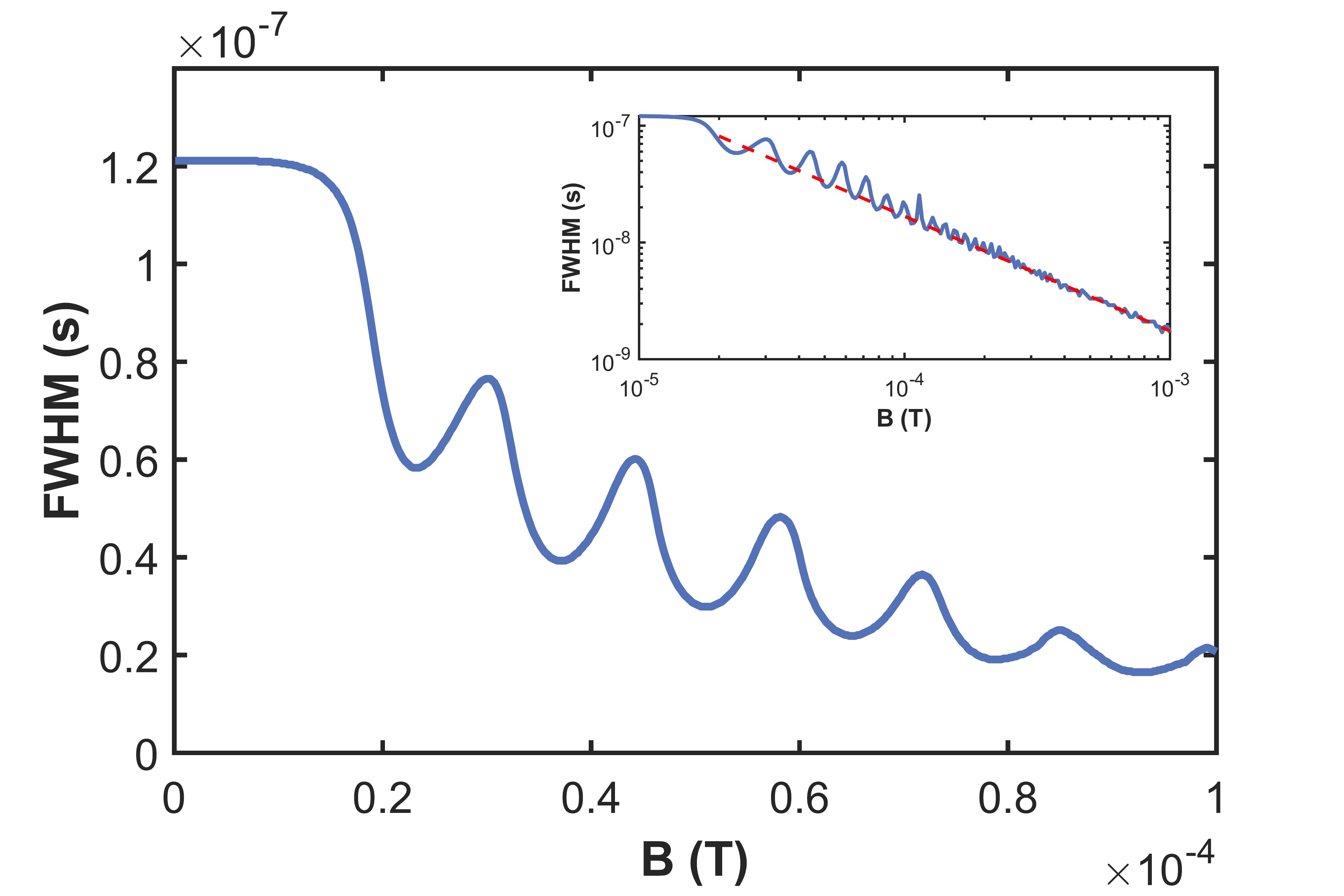}
\caption{\textbf{The linewidth of the correlation peak as a function of the magnetic field} \newline We plot the full-width at half-maximum (FWHM) of the normalized correlation peak ($t_d = 100$ $\mu$s) in the time domain as a function of the magnetic field. Here, we show how the linewidth of the correlation measurement changes due to higher harmonics as the magnetic field of the 500 kHz AC signal varies up to 100 $\mu$T (1 mT for inset), with $N_s$ = 10. In the inset, the slope of the red line obtained from a linear fit is found to be $-1.00 \pm 0.03$.}
\label{fig:frequency_resolution_linewidth}
\end{figure}

Not only dynamical decoupling pulse sequences itself can decouple a wider range of noise frequencies by using more $\pi$-pulses. Higher harmonics lead to a narrowing of the linewidth, and an improved frequency resolution for the same time $T$. A simple example is the change in linewidth of the peak at 100 $\mu$s in Fig.~\ref{fig:correlation_measurement_harmonics} for a 500 kHz AC magnetic field using a spin echo pulse sequence with $N_s=10$. The change in linewidth as a function of magnetic field strength can be seen in Fig.~\ref{fig:frequency_resolution_linewidth}. In particular, the inset of Fig.~\ref{fig:frequency_resolution_linewidth} reveals, via linear fitting, that the frequency resolution extracted from the linewidth of the correlation measurement scales not only as $\sim 1/T$, but also as $\sim 1/B$. This indicates that increasing the magnetic field amplitude, which induces higher harmonics, allows for achieving even higher frequency resolution. Note that similar results can be obtained by increasing the number of $\pi$-pulses instead of changing the magnetic field amplitude, since $\varphi_{\max} \propto B N$.

\section{Signal-to-Noise Ratio}
\subsection{Correlation Measurement}
To evaluate the performance of our protocol we investigate its achievable signal-to-noise ratio. 

As described above, we collect $N_\phi\times N_s$ measurements that we assume follow a Poisson distribution as the measurements are photon-shot-noise limited. 
In order to obtain the correlation from the signal variance as in Eq.~(\ref{n_sensor correlations}) we need to measure over a finite number of different phases, $N_\phi$. We obtain this by partitioning the total number of measurements into $N_\phi$ sets of $N_s$ measurements (in principle, the time delays between the $N_\phi$ sets can be different than $t_{d}$ introduced above). This results in a total of $N_s \times N_\phi$ measurements, each labeled as $S_{i,k}$. Then, the variance over this discrete set of measurements is 
\begin{equation}\label{eqn:SNR_corr0}
    \textrm{Var}(\mathbb{P})\equiv\text{Var}\left( \frac{1}{N_s} \sum_{i=1}^{N_s} S_i \right) = \frac{1}{N_\phi} \sum_{k=1}^{N_\phi} \left( \frac{1}{N_s} \sum_{i=1}^{N_s} S_{i,k} \right)^2 - \left( \frac{1}{N_\phi} \sum_{k=1}^{N_\phi} \frac{1}{N_s} \sum_{i=1}^{N_s} S_{i,k} \right)^2.
\end{equation}
Here $S_{i,k}$ represents the measurement at the $i$-th sensing delay (ranging from 1 to $N_s$) and the $k$-th phase (ranging from 1 to $N_\phi$), see Fig. 1. 
We then take as our  signal  the expected value of the variance, $\mathbb{E}[\textrm{Var}(\mathbb{P})]$, with its uncertainty set by the standard deviation of the variance, $\sqrt{\textrm{Var}[\textrm{Var}(\mathbb{P})]}$.
\subsubsection{Signal}
The measurements of $S_{i,k}$ follow a Poisson distribution for the  number of detected photons  with a parameter $\lambda_{i,k}$ (that we can take to be proportional to the population at the measurement $i,k$, $\lambda_{j,k}=\lambda P_{\ket0}^{i,k}$, Eq.~(\ref{eq:populationik}).) The expectation value  $\mathbb{E}[S_{i,k} S_{j,k}]$ depends on whether $i = j$, i.e., whether the measurements are independent. Thus, by transforming Eq.~(\ref{eqn:SNR_corr0}), we obtain
\begin{equation}\label{eqn:SNR_corr1}
\begin{split}
    \\&\text{Var}\left( \frac{1}{N_s} \sum_{i=1}^{N_s} S_i \right) = \frac{1}{N_\phi N_s^2} \sum_{k=1}^{N_\phi} \sum_{i=1}^{N_s} S_{i,k}^2 + \frac{1}{N_\phi N_s^2} \sum_{k=1}^{N_\phi} \sum_{i=1}^{N_s} \sum_{j \neq i}^{N_s} S_{i,k} S_{j,k} \\- \frac{1}{N_\phi^2 N_s^2}& \sum_{k=1}^{N_\phi} \sum_{i=1}^{N_s} S_{i,k}^2 - \frac{1}{N_\phi^2 N_s^2} \sum_{k=1}^{N_\phi} \sum_{i=1}^{N_s} \sum_{j \neq i}^{N_s} S_{i,k} S_{j,k} - \frac{1}{N_\phi^2 N_s^2} \sum_{k=1}^{N_\phi} \sum_{h \neq k}^{N_\phi} \sum_{i=1}^{N_s} \sum_{j=1}^{N_s} S_{i,k} S_{j,h}.
\end{split}
\end{equation}

Using the expectation values from the Poisson distribution as shown in Eq.~(\ref{eqn:SNR_corr6}), the signal of the correlation measurement can be expressed as
\begin{equation}\label{eqn:SNR_corr2}
\begin{split}
    \\\mathbb{E}\left[\text{Var}\left( \frac{1}{N_s} \sum_{i=1}^{N_s} S_i \right)\right] =& \frac{N_\phi-1}{N_\phi^2 N_s^2} \sum_{k=1}^{N_\phi} \sum_{i=1}^{N_s}  \lambda_{i,k} + \frac{1}{N_\phi N_s^2} \sum_{k=1}^{N_\phi} \sum_{i=1}^{N_s} \sum_{j=1}^{N_s} \lambda_{i,k} \lambda_{j,k}- \frac{1}{N_\phi^2 N_s^2} \sum_{k=1}^{N_\phi} \sum_{h=1}^{N_\phi} \sum_{i=1}^{N_s} \sum_{j=1}^{N_s} \lambda_{i,k} \lambda_{j,h}.
\end{split}
\end{equation}

If all of the $N_s$ measurements are identically distributed, i.e., $\lambda_{i,k} = \lambda_{j,k} = \lambda_{k}$, the signal can be written as
\begin{equation}\label{eqn:SNR_corr3}
    \mathbb{E}\left[\text{Var}\left( \frac{1}{N_s} \sum_{i=1}^{N_s} S_i \right)\right] = \frac{N_\phi - 1}{N_\phi^2 N_s} \sum_{k=1}^{N_\phi} \lambda_k + \frac{1}{N_\phi N_s} \sum_{k=1}^{N_\phi} \lambda_k^2 
  - \left(\frac{1}{N_\phi} \sum_{k=1}^{N_\phi}  \lambda_k \right)^2.
\end{equation}

\subsubsection{Noise}
The noise of the variance can be obtained through the calculation of the variance of the variance
\begin{equation}\label{eqn:SNR_corr4}
    \text{Var}\left[ \text{Var} \left( \frac{1}{N_s} \sum_{i=1}^{N_s} S_i \right) \right] = \mathbb{E}\left[ \text{Var} \left( \frac{1}{N_s} \sum_{i=1}^{N_s} S_i \right)^2 \right] - \mathbb{E}\left[ \text{Var} \left( \frac{1}{N_s} \sum_{i=1}^{N_s} S_i \right) \right]^2.
\end{equation}

The second moment $\mathbb{E}\left[ \text{Var} \left( \frac{1}{N_s} \sum_{i=1}^{N_s} S_i \right)^2 \right]$ is calculated as follows
\begin{equation}\label{eqn:SNR_corr5}
\begin{split}    
    \\&\text{Var}\left( \frac{1}{N_s} \sum_{i=1}^{N_s} S_i \right)^2 = \frac{1}{N_\phi^2 N_s^4} \sum_{k=1}^{N_\phi} \sum_{i=1}^{N_s} \sum_{j=1}^{N_s} \sum_{k' = 1}^{N_\phi} \sum_{i' = 1}^{N_s} \sum_{j' = 1}^{N_s} S_{i,k} S_{j,k} S_{i',k'} S_{j',k'}\\ - \frac{2}{N_\phi^3 N_s^4} \sum_{k=1}^{N_\phi} \sum_{h=1}^{N_\phi} \sum_{i=1}^{N_s} \sum_{j=1}^{N_s}& \sum_{k' = 1}^{N_\phi} \sum_{i' = 1}^{N_s} \sum_{j' = 1}^{N_s} S_{i,k} S_{j,h} S_{i',k'} S_{j',k'} + \frac{1}{N_\phi^4 N_s^4} \sum_{k=1}^{N_\phi} \sum_{h=1}^{N_\phi} \sum_{i=1}^{N_s} \sum_{j=1}^{N_s} \sum_{k' = 1}^{N_\phi} \sum_{h' = 1}^{N_\phi} \sum_{i' = 1}^{N_s} \sum_{j' = 1}^{N_s} S_{i,k} S_{j,h} S_{i',k'} S_{j',h'}
\end{split}
\end{equation}

Since the measurements follow a Poisson distribution, the expected values for the different powers of the distribution are

\begin{equation}\label{eqn:SNR_corr6}
\begin{split}    
    \\\mathbb{E}[x] =& \lambda \\ \mathbb{E}[x^2] = \lambda& (\lambda + 1) \\ \mathbb{E}[x^3] = \lambda (\lambda^2& + 3\lambda + 1) \\ \mathbb{E}[x^4] = \lambda (\lambda^3 + 6&\lambda^2 + 7\lambda + 1)
\end{split}
\end{equation}

As a result of the calculation, the variance of $ \text{Var} \left( \frac{1}{N_s} \sum_{i=1}^{N_s} S_i \right) $ for a Poisson distribution follows
\begin{equation}\label{eqn:SNR_corr7}
\begin{split}    
    \\\text{Var}&\left[\text{Var}\left( \frac{1}{N_s} \sum_{i=1}^{N_s} S_i \right)^2\right] = 
    \frac{(N_{\phi}-1)^2}{N_{\phi}^{4}N_{s}^{4}}\sum_{k=1}^{N_{\phi}}\sum_{i=1}^{N_{s}} \lambda_{i,k}
    +\frac{6N_{\phi}^2-8N_{\phi}}{N_{\phi}^{4}N_{s}^{4}}\sum_{k=1}^{N_{\phi}}\sum_{i=1}^{N_{s}}\sum_{j=1}^{N_{s}} \lambda_{i,k}\lambda_{j,k}
    \\
    &+\frac{4N_{\phi}^2}{N_{\phi}^{4}N_{s}^{4}}\sum_{k=1}^{N_{\phi}}\sum_{i=1}^{N_{s}}\sum_{j=1}^{N_{s}}\sum_{j'=1}^{N_{s}} \lambda_{i,k}\lambda_{j,k}\lambda_{j',k}
    +\frac{6-4N_{\phi}}{N_{\phi}^{4}N_{s}^{4}}\sum_{k=1}^{N_{\phi}}\sum_{h=1}^{N_{\phi}}\sum_{i=1}^{N_{s}}\sum_{j=1}^{N_{s}} \lambda_{i,k}\lambda_{j,h}
    \\
    -&\frac{8N_{\phi}}{N_{\phi}^{4}N_{s}^{4}}\sum_{k=1}^{N_{\phi}}\sum_{h=1}^{N_{\phi}}\sum_{i=1}^{N_{s}}\sum_{j=1}^{N_{s}}\sum_{j'=1}^{N_{s}} \lambda_{i,k}\lambda_{j',k}\lambda_{j,h}
    +\frac{4}{N_{\phi}^{4}N_{s}^{4}}\sum_{k=1}^{N_{\phi}}\sum_{i=1}^{N_{s}}\sum_{h=1}^{N_{\phi}}\sum_{h'=1}^{N_{\phi}}\sum_{j=1}^{N_{s}}\sum_{i'=1}^{N_{s}} \lambda_{i,k}\lambda_{j,h}\lambda_{i',h'}
\end{split}
\end{equation}

\subsubsection{SNR}
In the case of correlation measurement, the signal corresponds to the case when correlations occur. Therefore, the variance for the correlated case is given by
\begin{equation}\label{eqn:SNR_corr8}
\begin{split}    
    \\ \text{Var}&\left[\text{Var}\left( \frac{1}{N_s} \sum_{i=1}^{N_s} S_i \right)^2\right] = \frac{(N_{\phi} - 1)^2}{N_{\phi}^4 N_s^3} \sum_{k=1}^{N_{\phi}} \lambda_k + \frac{6N_{\phi}^2 - 8N_{\phi}}{N_{\phi}^4 N_s^2} \sum_{k=1}^{N_{\phi}} \lambda_k^2 + \frac{4N_{\phi}^2}{N_{\phi}^4 N_s} \sum_{k=1}^{N_{\phi}} \lambda_k^3 \\&+ \frac{6 - 4N_{\phi}}{N_{\phi}^4 N_s^2} \sum_{k=1}^{N_{\phi}} \sum_{h=1}^{N_{\phi}} \lambda_k \lambda_h - \frac{8N_{\phi}}{N_{\phi}^4 N_s} \sum_{k=1}^{N_{\phi}} \sum_{h=1}^{N_{\phi}} \lambda_k^2 \lambda_h + \frac{4}{N_{\phi}^4 N_s} \sum_{k=1}^{N_{\phi}} \sum_{k' = 1}^{N_{\phi}} \sum_{h' = 1}^{N_{\phi}} \lambda_k \lambda_{k'} \lambda_{h'}.
\end{split}
\end{equation}

Subsequently, the Signal-to-Noise Ratio (SNR) can be easily obtained as $\text{SNR} = \frac{\mu}{\sigma}$, where $\mu$ and $\sigma$ are the mean and standard deviation, respectively.

\subsection{Synchronized Readout Measurement}

\subsubsection{Signal}
In order to compare our protocol with existing ones, in particular synchronized readout, we assume to perform  $N=N_\phi \times N_s$ readouts $S_i$ of the AC signal. In the synchronized readout protocols introduced in Refs.~\cite{boss2017quantum,aslam2017nanoscale}, 
the experimental signal is analyzed via its periodogram, 
\begin{equation}\label{eqn:SNR_sync1}
    |y(\nu)|^2 = \sum_{k=0}^{N-1} \sum_{h=0}^{N-1} S_k S_h e^{-i \frac{2\pi\nu (h-k)}{N}}.
\end{equation}

Since $S_k$ and $S_h$ follow a Poisson distribution, the expectation value of the signal $\mathbb{E}[|y(\nu)|^2]$ becomes
\begin{equation}\label{eqn:SNR_sync2}
    \mathbb{E}[|y(\nu)|^2] = \sum_{k=0}^{N-1} \sum_{h=0}^{N-1} \lambda_k \lambda_h e^{-i \frac{2\pi\nu (h-k)}{N}} + \sum_{k=0}^{N-1} \lambda_k.
\end{equation}

\subsubsection{Noise}
The variance of the synchronized readout periodogram can be computed by first evaluating $|y(\nu)|^4$: 
\begin{equation}\label{eqn:SNR_sync3}
    |y(\nu)|^4 = \sum_{k=0}^{N-1} \sum_{h=0}^{N-1} \sum_{k' = 0}^{N-1} \sum_{h' = 0}^{N-1} S_k S_h S_{k'} S_{h'} e^{-i \frac{2\pi\nu (h-k)}{N}} e^{-i \frac{2\pi\nu (h'-k')}{N}}.
\end{equation}

By utilizing the result from $\mathbb{E}[|y(\nu)|^2]^2$, the variance is 
\begin{equation}\label{eqn:SNR_sync4}
\begin{split}
    \\&\text{Var} [|y(\nu)|^2] = \mathbb{E}[|y(\nu)|^4] - \mathbb{E}[|y(\nu)|^2]^2\\=\sum_{k=0}^{N-1}\lambda_{k}+4\sum_{k=0}^{N-1}\sum_{k'=0}^{N-1}&\lambda_{k}\lambda_{k'}e^{-i\frac{2\pi\nu(k'-k)}{N}}+\sum_{k=0}^{N-1}\sum_{k'=0}^{N-1}\lambda_{k}\lambda_{k'}+\sum_{k=0}^{N-1}\sum_{h=0}^{N-1}\lambda_{k}\lambda_{h}e^{-2i\frac{2\pi\nu(h-k)}{N}}\\+2\sum_{k=0}^{N-1}\sum_{h=0}^{N-1}\sum_{h'=0}^{N-1}\lambda_{k}\lambda_{h}&\lambda_{h'}e^{-i\frac{2\pi\nu(h-k)}{N}}e^{-i\frac{2\pi\nu(h'-k)}{N}}+2\sum_{k=0}^{N-1}\sum_{h=0}^{N-1}\sum_{h'=0}^{N-1}\lambda_{k}\lambda_{h}\lambda_{h'}e^{-i\frac{2\pi\nu(h-h')}{N}}.
\end{split}
\end{equation}

\subsubsection{SNR}
Assuming that the synchronized readouts are done at time intervals $t_{d}$, given the signal in Eq.~(\ref{echopopulation2}) with $\phi_0=i\omega t_{d}$, we expect the main peak of the DFT to be at the first harmonic, $\nu=2N_s$. However, for particular values of the AC field amplitude and a given decoupling pulse sequence, other harmonics might increase~\cite{mizuno2020simultaneous}. Given the complexity of the data analysis in that case,  here we focus on $|y(\nu=2N_s)|^2$ only.

\subsection{Numerical Simulations}

To compare the performance of the two measurement methods, correlation measurement and synchronized readout, which provide arbitrary frequency resolution, we first calculated the SNR. To verify the accuracy of the calculations, a numerical simulation was conducted. In both cases, we assumed an ensemble of NV centers with a sufficiently high density to neglect quantum projection noise, and simplified the scenario by assuming that for each qubit, one photon is detected when it is in the state $|0\rangle$ and zero photons are detected when it is in the state $|-1\rangle$. Additionally, we assumed each NV center had a collection efficiency $\eta = 0.0001$.

The expected number of photons to be observed depends on the phase of the AC magnetic field as described by Eq.~\ref{echopopulation2}. Therefore, the number of photons for each NV center at each measurement step is given by $\lambda_{i,k} = P_{|0\rangle,i,k} \cdot N_{\text{NV}} \cdot \eta$,
where $N_{\text{NV}}$ is the number of NV centers in the sensing volume. Assuming a random initial phase at the start of measurements and continuous application of the AC magnetic field, $\lambda_{i,k}$ can be determined based on the measurement interval. These values of $\lambda_{i,k}$ can then be used in the previously calculated SNR formulas for correlation measurement and synchronized readout to generate the corresponding plots.

For the numerical simulation, each $\lambda_{i,k}$ is treated as a Poisson-distributed random number representing the expected number of photons. After generating these Poisson random variables for the entire measurement process, the signal can be obtained by substituting them into Eq.~(\ref{eqn:SNR_corr3}) for the correlation measurement, while for synchronized readout, the discrete Fourier transform (DFT) is applied to extract the peak. Note that we used squared values as shown in previous works on synchronized readout methods\cite{boss2017quantum,aslam2017nanoscale}. By repeating this procedure 1000 times, we obtain 1000 signal values, from which the mean value is taken as the signal, and the standard deviation as the noise, from which the SNR is calculated. This process is shown in Fig. 2.d of the main manuscript. Using this approach, we can observe how the SNR changes with the amplitude of the AC magnetic field.

In the synchronized readout method, a region is observed where the signal nearly disappears in the SNR plot, a phenomenon that has already been reported in prior works\cite{mizuno2020simultaneous}. Since sensing typically focuses on detecting tiny fields, this limitation may not be critical. However, when the signal appears weak in our measurements, efforts are made to increase sensitivity, such as increasing the number of $\pi$ pulses in pulse sequences like XY8. The number of $\pi$ pulses can be related to the phase accumulated by the signal, as described by Eq.~(\ref{eq:CP}). If too many $\pi$ pulses are added, it may lead to a region where measurements are not possible. Therefore, when using synchronized readout, knowledge of the amplitude of the B-field in advance is necessary to avoid this issue.

\begin{figure}[t]
\centering
\includegraphics[width=0.7\linewidth]{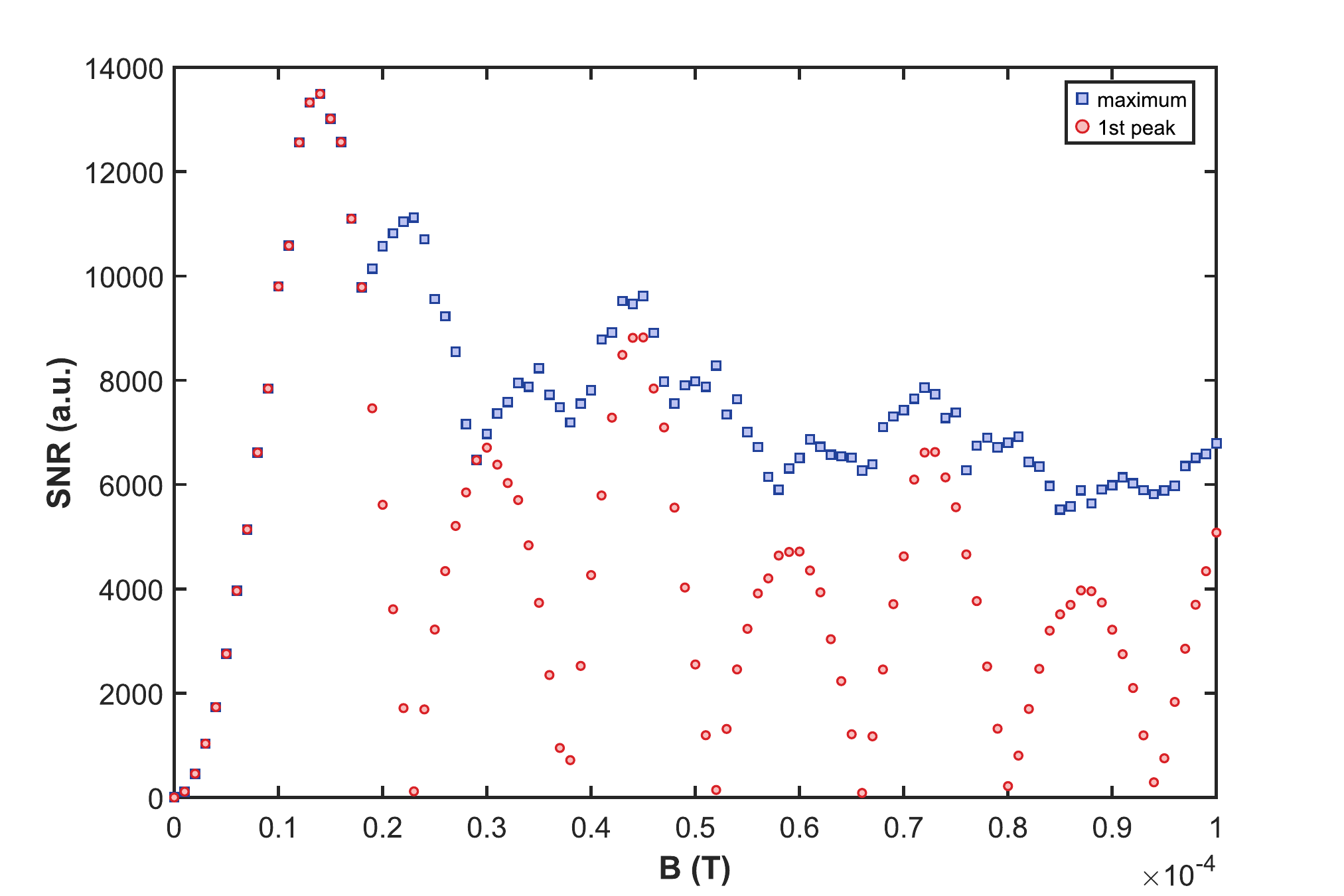}
\caption{\textbf{Simulation of SNR variation using 1st peak and maximum value in DFT as a function of the B-field amplitude in the synchronized readout}\newline In the simulation, we assume the observation of an ensemble NV layer with a diameter of 40 $\mu$m and a thickness of 30 $\mu$m, with a concentration of 1.25 ppm. The collection efficiency of each NV center is assumed to be 0.0001. We generated a sequence of Poisson random variables for the $\lambda P_{\ket{0}}$, corresponding to 10,000 measurements (with 1,000 different phases of the 500 kHz AC magnetic field, denoted by $N_{\phi}=1000$, and 10 consecutive periodic measurements, denoted by $N_{s}=10$). A DFT was then performed. The process was repeated 1,000 times to extract the 1st peak at $\nu=20$ (red circle) and the maximum (blue square) of the absolute square of DFT. From these results, the mean and standard deviations of the random variables were calculated to obtain the SNR.}
\label{fig:simulation_3}
\end{figure}

In synchronized readout, an alternative approach could involve using peaks other than the first peak when the signal is undetectable. However, this method is not ideal for two reasons: first, the "relative frequency difference" measured by synchronized readout could be misinterpreted; second, because the oscillation frequency does not transition fully to a different frequency, the various peaks will share the signal. In our simulation, we suggested searching for the maximum point instead of the first peak, which avoids the region where the signal vanishes. As seen in Fig.~\ref{fig:simulation_3}, this approach does not result in any region where the signal is zero but still provides a lower SNR than correlation measurement.

\begin{figure}[t]
\centering
\includegraphics[width=0.7\linewidth]{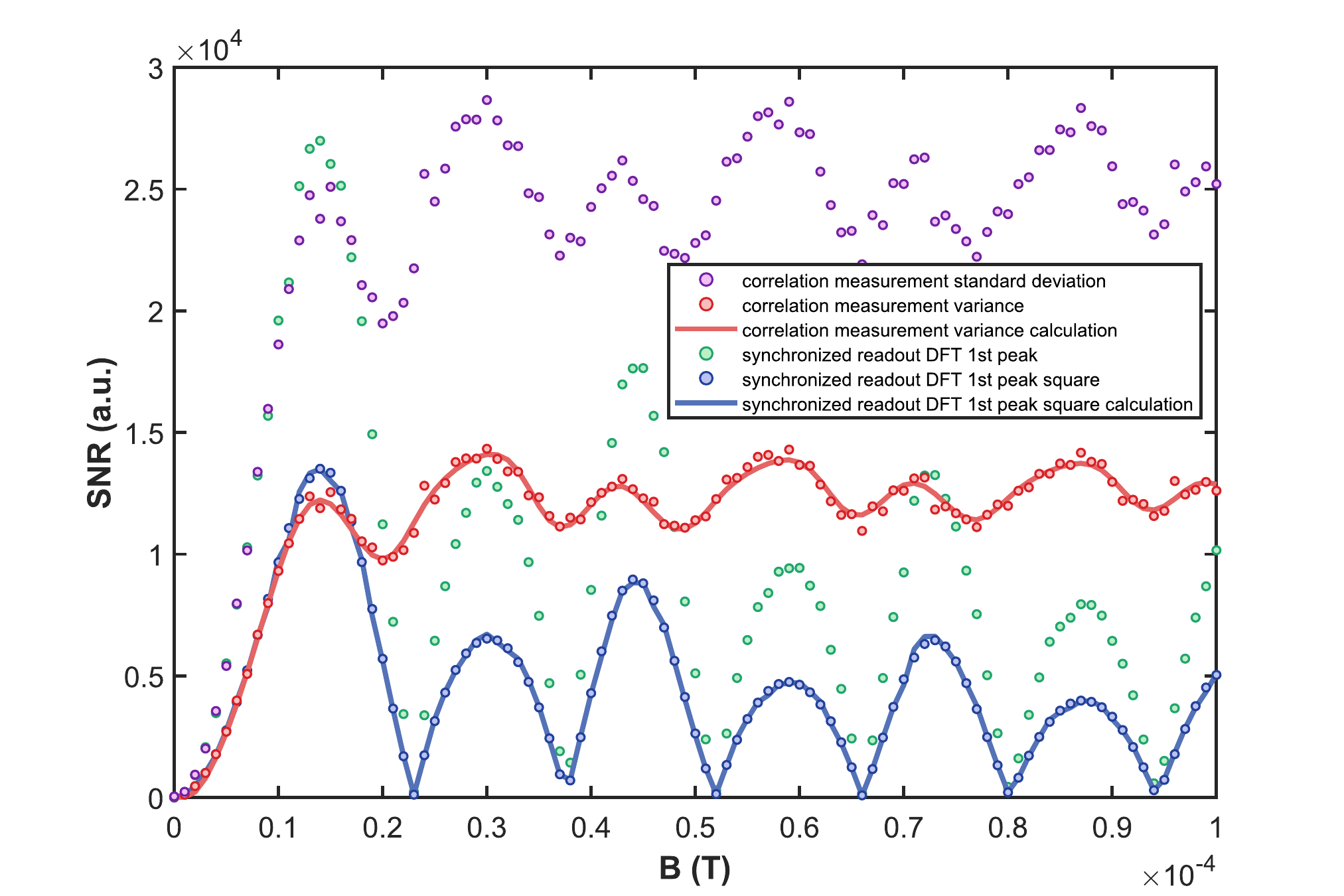}
\caption{\textbf{SNR variation as a function of the B-field amplitude in the synchronized readout and correlation measurement methods, both with and without normalization} \newline We used the same parameters of NV samples in Fig.~\ref{fig:simulation_3} for the simulations. We generated a sequence of Poisson random variables for the $\lambda P_{\ket{0}}$, corresponding to 10,000 measurements (with 1,000 different phases of the 500 kHz AC magnetic field, denoted by $N_{\phi}=1000$, and 10 consecutive periodic measurements, denoted by $N_s = 10$). This process was repeated 1,000 times to obtain the SNR for the synchronized readout and correlation measurements, as same as Fig.~\ref{fig:simulation_3}. For the SNR of the synchronized readout, the absolute square (blue) and the normalized absolute value (green) of the 1st peak in the DFT are represented. In the SNR of the correlation measurement, the variance (red), and the normalized standard deviations (purple) are used.}
\label{fig:simulation_2}
\end{figure}

Finally, for the simulation, we used standard deviations instead of variances in the correlation measurement and did not square the values in synchronized readout, allowing the use of normalized values. Note that we used standard deviations instead of variance in our experiment for normalization. Since there is no known method for calculating the square root term in a Poisson random distribution, we calculated the variance instead of the standard deviation. However, it is still possible to simulate and compare the SNR for normalized and non-normalized cases, as demonstrated in Fig.~\ref{fig:simulation_2}. Both methods showed an approximately twofold increase in SNR when normalized, which can be intuitively understood through the following simple calculation.

Let $x$ represent our measurement and $\lambda$ be the expected number of photons. After measuring $x$, the value of $x^2$ can be obtained by squaring $x$. The SNR for $x$ and $x^2$ are calculated as follows:
\begin{equation}
    SNR(x)=\frac{E[x]}{\sqrt{E[x^{2}]-E[x]^{2}}}=\sqrt{\lambda},
\end{equation}
\begin{equation}
    SNR(x^{2})=\frac{E[x^{2}]}{\sqrt{E[x^{4}]-E[x^{2}]^{2}}}=\frac{\lambda(\lambda+1)}{\sqrt{\lambda(4\lambda^{2}+6\lambda+1)}}\approx\frac{\sqrt{\lambda}}{2}.
\end{equation}
Thus, without normalization, the SNR decreases by approximately a factor of 2 in Poisson-limited measurements, which can be understood through these calculations.

\section{Methods}
\subsection{Setup and Control}

In our experiment, we used CVD-grown diamond (Element Six, with a $^{14}$N concentration of $\sim$10 ppm and 99.999\% purified $^{12}$C). A field of view of approximately 40 $\mu$m in diameter was collected by a photodiode (PDA36A). To explore different experimental directions, we applied a magnetic field with components of 2.1, 10.6, and 19.5 Gauss to ensure that the resonance frequencies of NV centers along different axes were not degenerate.

For homogeneous spin manipulation in the measurement region, RF and AC magnetic fields were applied using a two-layer circular loop PCB. To generate a spin echo pulse sequence, we utilized a pulse generator (Spin Core, PulseBlaster), and the RF signal was generated by mixing a 100 MHz analog signal from an arbitrary waveform generator (AWG) with the RF local oscillation from an RF generator (Windfreak, SynthNV).

When sweeping the $t_{d}$, the first parameter to be swept was set using the PulseBlaster, and additional time delays were swept with the AWG. For example, when sweeping from 400 $\mu$s to 402 $\mu$s, a large $t_{d}$ of 400 $\mu$s was set with the PulseBlaster, and the additional time delay from 0 to 2 $\mu$s was generated using the AWG. It should be noted that, in principle, the PulseBlaster alone would suffice; however, due to the higher sample rate of the AWG compared to the PulseBlaster, finer $t_{d}$ measurements could be achieved, and thus this method was employed for our measurements.

In our experimental setup, to mitigate the effects of large laser noise, we performed reference measurements for each single measurement and normalized the results accordingly. However, in principle, our protocol can also be used without reference measurements.

\subsection{Experimental Details}
\subsubsection{Random phases}\label{sec:randomphase1}
As introduced in Sec.~\ref{finite_avg}, experimentally implementing the measurement of a randomly distributed initial phase using the method outlined in Eq.~(\ref{finite_phases}) can be easily achieved if the approximate frequency of the AC magnetic field is known. For instance, in our experiment, when measuring a 500 kHz magnetic field as shown in Fig. 1.b, the AC magnetic field will repeat the same phase every 2 $\mu$s. Therefore, if we want to divide the measurement into $N_\phi$ phases, we can measure at intervals of $m \cdot 2 \mu$s + $2 \mu$s/$N_\phi$, where $m$ is a natural number. As $N_\phi$ increases, Eq.~(\ref{finite_phases}) will converge to Eq.~(\ref{eq:mean}), regardless of the initial phase $\phi_0$.
This method offers a significant advantage because, even if the initial phase changes during repeated measurements, the same sequential measurement process can yield consistent results.

\subsubsection{Correlation measurement}

The population of a single measurement is given by Eq.~(\ref{echopopulation2}). If a measurement is performed after a time delay $t_{d}$, an additional phase of $\omega t_{d}$ will be added to $\phi_0$. While the frequency of the AC magnetic field is $\omega$, the population follows a $\cos^2$ dependence, resulting in a frequency of $2\omega$. In Eq.~(\ref{eq:nmeasurements0}), when $\omega t_{d} = m \pi$, where $m$ is a natural number, the results will depend on the initial phase $\phi_0$, and measurements performed with different $\phi_0$ will yield the standard deviation due to variations in $\phi_0$ as
\begin{equation}
    \mathbb{P}=\frac1{N_s}\sum_{i=0}^{N_s-1} P_{\ket0 , i}=\cos^{2}{[2\vartheta\cos{\left(\frac{\omega\tau}{2}+\phi_{0}\right)}]}.
\end{equation}
Note that, using the method described in Sec.~\ref{sec:randomphase1}, measurements can be accumulated for different values of $\phi_0$, so the initial value of $\phi_0$ does not matter. As a result, even when the initial phase $\phi_0$ differs during repetition due to insufficient SNR in a single measurement, the expected measurement result remains the same, and synchronization is not required.

On the other hand, when the time delay $\omega t_{d}$ satisfies the condition $\omega t_{d} = m \pi + \frac{k \pi}{N_s}$, where $k = 1, 2, 3, \dots, N_s-1$, measurements will be made across a range of phases as discussed in Sec.~\ref{sec:randomphase1} for random phases, and the results will follow a Bessel function, regardless of the initial phase $\phi_0$
\begin{equation}
    \mathbb{P}=\frac1{N_s}\sum_{i=0}^{N_s-1} P_{\ket0 , i}\simeq\frac{1}{2}+\frac{1}{2}J_{0}[4\vartheta].
\end{equation}
However, when the time delay satisfies $\omega t_{d} = (m + 1) \pi$, the results will again depend on the initial phase $\phi_0$. In this case, the variance resulting from changes in $\phi_0$ will be observed, leading to a correlated condition. Therefore, when performing a correlation measurement using $N_s$ measurements, there will be $N_s - 1$ uncorrelated time delays between these two correlated conditions. The interval between each of these uncorrelated time delays is $\frac{\pi}{\omega N_s}$ setting the frequency resolution of our scheme (for small B fields).

Let us assume that $\omega t_{d}$ satisfies the correlated conditions in the correlation measurement. If the frequency is detuned by an additional $\Delta \omega$, the frequency becomes uncorrelated. In this case, one signal will be measured while the other will not, and the signals can be distinguished. The nearest uncorrelated time for the correlated condition will occur when $\Delta \omega t_{d} = \pi / N_s$. The total measurement time required for this process is $t_{d}N_s$, and thus the frequency resolution $\Delta \omega$ is inversely proportional to the total measurement time, $t_{d}N_s$.

In experiments, to measure the correlation between $N_s$ measurements, the time delay $t_{d}$ is swept across $N_\phi$ different phases to obtain the variance. For each measurement, the $i$-th measurement and the $k$-th phase out of $N_\phi$ will satisfy the following relation
\begin{equation}
    P_{|0\rangle}^{i, k} = \cos^2 \left[ 2\vartheta \cos\left(\frac{\omega \tau}{2} + \phi_0 + i \omega t_{d} + \frac{2\pi k}{N_\phi}\right) \right].
    \label{eq:populationik}
\end{equation}

For example, in the correlation measurement of five measurements with a sweep of $t_{d}$ from 400 $\mu$s to 404 $\mu$s as shown in Fig. 1.c, measurements are repeated for 20 different phases. For this, the period of the 500 kHz magnetic field, which is 2 $\mu$s, is divided by 20 to achieve a 100 ns detuned time between measurements. After completing the $N_s$ measurements, the second measurement is performed 100 ns after the first, following a 2 ms time interval. In this way, measurements for all 20 $N_\phi$ phases can be performed, allowing for correlation between the measurements.

To achieve arbitrary frequency resolution, the time delay $t_{d}$ increases, and as shown in Fig. 2.a, dead time occurs during the execution of $N_s$ measurements. During this dead time, measurements with a 100 ns time delay can still be conducted. Note that the time interval between the $k$-th phase measurements remains constant at $t_{d}$.

\subsection{Data Processing} 
If we perform correlation measurements using $N_{s}$ distinct measurements, and measure $N_{\phi}$ different phases, the measurement result will be

\begin{equation}
    \textrm{Var}\left(\frac{1}{N_{s}}\sum_{i=1}^{N_{s}} S_{i}\right)=\frac{1}{N_{s}^{2}}\sum_{i=1}^{N_{s}}\sum_{j=1}^{N_{s}}Cov\left(S_{i},S_{j}\right).
\end{equation}

Thus, the standard deviation will be

\begin{equation}
    \sigma=\frac{1}{N_{s}}\sqrt{\sum_{i=1}^{N_{s}}\sum_{j=1}^{N_{s}}Cov\left(S_{i},S_{j}\right)}.
\end{equation}

In experiments, we used a finite value of $N_{\phi}$. Therefore, the standard deviation of the signal will be

\begin{equation}
    \hat{\sigma}=\sqrt{\frac{1}{N_{\phi}}\sum_{k=1}^{N_{\phi}}\left(\frac{1}{N_{s}}\sum_{i=1}^{N_{s}} S_{i,k}\right)^{2}-\left(\frac{1}{N_{\phi}}\sum_{k=1}^{N_{\phi}}\frac{1}{N_{s}}\sum_{i=1}^{N_{s}} S_{i,k}\right)^{2}}.
\end{equation}

We repeated this measurement $N_{r}$ times. Thus, the standard deviation of the estimate $\hat{\sigma}$ is

\begin{equation}
    \sigma_{\hat{\sigma}}=\sqrt{\frac{1}{N_{r}-1}\sum_{t=1}^{N_{r}}\left(\hat{\sigma}_{t}-\bar{\hat{\sigma}}\right)^{2}},
\end{equation}
where $\bar{\hat{\sigma}}=\frac{1}{N_{r}}\sum_{t=1}^{N_{r}}\hat{\sigma}_{t}$.

\putbib[bibliography]
\end{bibunit}

\end{document}